\documentclass{JHEP3} 

\usepackage{multicol,bbm}
\usepackage{graphicx, epsfig, subfigure}

\newcommand\fverb{\setbox\fverbbox=\hbox\bgroup\verb}
\newcommand\fverbdo{\egroup\medskip\noindent%
            \fbox{\unhbox\fverbbox}\ }
\newcommand\fverbit{\egroup\item[\fbox{\unhbox\fverbbox}]}
\newbox\fverbbox

\newcommand{\be}{\begin{equation}}
\newcommand{\ee}{\end{equation}}
\newcommand{\bea}{\begin{eqnarray}}
\newcommand{\eea}{\end{eqnarray}}
\newcommand{\ba}{\begin{array}}
\newcommand{\ea}{\end{array}}

\def\a{\alpha}

\def\o{\omega}

\def\th{\theta}
\def\r{\rho}

\def\t{\tau}

\def\O{\Omega}

\title{Holographic  Instanton Liquid and chiral transition}

\author{Bogeun Gwak$^{a,b}$, Minkyoo Kim$^{a,b}$, Bum-Hoon Lee$^{a,b}$, Yunseok Seo$^{b}$ and Sang-Jin Sin$^{c}$
\\
$^{a}$Department of Physics, Sogang University, Seoul 121-742, Korea
\\
$^{b}$Center for Quantum Spacetime, Sogang University, Seoul 121-742, Korea
\\
$^{c}$Department of Physics, Hanyang University, Seoul 133-791, Korea
\\
Email: \email{rasenis@sogang.ac.kr, mkim80@sogang.ac.kr, bhl@sogang.ac.kr, yseo@sogang.ac.kr, sjs@hanyang.ac.kr}}

\abstract{We study the phase diagram of {\it black} D3 geometry with uniformly distributed D-instanton charge using the probe D7 brane.   
In the presence of uniform D-instanton charges,  quarks can be confined
although gluons are not,   because 
baryon vertices are allowed  due to the net repulsive force on the on the probe D-branes.  
Since there is no scale in the geometry itself apart from the horizon size, 
there is no Hawking-Page transition. 
As a consequence, the D7 brane embedding can encode the effect of the 
the finite temperature as well as finite baryon density even for low temperature. 
The probe D-brane embedding, however, undergoes 
a chiral phase transition according to the temperature and density parameter. 
 We studied such phase transitions  and calculated the constituent quark mass, chiral condensation and the binding energy of baryons as function of the density. The baryon vertex melting is identified as the quark deconfinement. We draw phase diagram according to these transitions. }

\begin{document}

\section{Introduction}

One of the difficulty in holographic QCD is to discuss the temperature dependence of physical observables in the hadron phase.  
The problem is the presence of the Hawking-Page transition  as the deconfinement/confinement transition \cite{witten2}: background metric  describing the low temperature phase does not contain  any temperature parameter. 
Describing the hadrons in finite temperature 
requests  having a black hole metric, in which gluons are always deconfined. 
Therefore we are lead to the question whether quark can be confined while 
gluons are deconfined.  
In fact the degree of freedom of two species are   different  in large $N_c$ order 
and the confinement of gluons and that of quarks should be separate phenomena at least in large $N_c$ limit. 
However, in most of the geometric background  with well defined Hawking temperature,  
a baryon vertex \cite{Witten:baryon} is not allowed as a finite energy solution\cite{callan,Seo:2008qc},  because black D-brane has net attractive gravity. 
 
To have a phase transition, we need  a scale other than the temperature.   
In  QCD, scale invariance is broken by the chiral and gluon condensation, and it is also known that 
both of them are induced by the presence of the instanton.   
 There has been huge activity to utilize the instanton background in gauge theory and 
 (quark) dynamics in the instanton background has been practiced by many  
the topic directly relevant to our paper is the instanton gas/liquid  model, where 
mass generation and the chiral symmetry breaking was discussed using the anomaly and fermion zero modes.  For a review, see the review article by   Schafer and Shuryak \cite{Schafer:1996wv} and references therein.
   
It is also well known that the  D-instanton is the gravity dual to the Yang-Mill instanton.
Therefore, it is very natural to try to use a background dual to the instanton gas. In fact, a background 
 which is dual to the unformly distributed D-instanton over D3 was suggested by  Hong Liu and Tseytlin
 \cite{Liu:1999fc}, where 
they also proposed, naturally,  that  the D3/D-instanton geometry is dual of the N=4 Yang Mill theory with constant gluon condensation with zero electric/magnetic gluon field, that is 
$<F_{\mu\nu}>=0, <{\rm Tr} F^{\mu\nu}F_{\mu\nu} >=cq\neq ,  0$ 
where $F$ is field strength of the gauge fields.
Such D3/D-instanton geometry contains non-trivial dilaton   giving  non-zero value of gluon condensation   $q$,  which was is   identified  precisely as the D-instanton density.  

While the chiral symmetry breaking was a natural consequence of the 
instanton background, the confinement,   was not   directly reachable  in the gauge theory instanton gas model.
One of the interesting consequence of the AdS/CFT is that
Liu and Tseytlin actually  have shown that the background has a confinement with a linear quark-antiquark potential.   Therefore in the gravity dual, both the chiral condensation and confinement turns out to be consequences of the presence of the D-instanton charges,  which produce  effectively net repulsive force on the probe D-branes and the probe Fundamental strings.  
 
The purpose of this paper is to  consider the finite temperature and density of the holographic dual of the D instanton gas/liquid.  
  For our purpose, we need to extend the geometry of Liu and Tseytlin to the finite temperature. 
  The necessary metric was   found by
  Ghoroku et al. \cite{Ghoroku}  in the context completely  unrelated to instanton gas model. 
   While most of the dilaton gravity solution is  singular \cite{gubser, park-sin} 
  such that  temperature is ill-defined, their solution  allows a Hawking temperature. 
   The geometry is quasi-confining, namely, it has following properties \cite{Ghoroku}: 
   (i) gluons are deconfined:   In QCD, one can consider the gluon propagator directly and state the (de)confinement. However, in hQCD, one can not  look at the 
gluon propagator directly since that is colored object whose treatment is beyond the gravity limit. 
Instead, what one looks at is the spectrum of the dilaton field  which is dual of the glueball operator.
IF that spectrum is well peaked or discrete, this is the signal of confinement. If the spectral function does not show any peak or  feature, it means they are all dis-integrated and gluons are in de-confined state. 
As a theorem,  the presence of the horizon does not allow any stable glueball spectrum, because
in AdS black hole,  quasi-normal mode spectrum  has large imaginary part.  
 (ii) the quark-anti quark potential  is Coulomb like for short distance, linear  for medium distance
   and screened for large enough distances. The turn-over distance depends on temperature and gluon condensation. 
(iii) Baryon vertex solution is allowed  in low enough temperature.

There is no geometric transition as temperature grows because (i) we do not have any compactified direction,  
(ii) We need to work out the thermodynamics of the bulk theory in Einstein frame, in which 
our geometry is the same as the black D3 brane, (iii) the action for the dilaton and axion cancel each other.
The difference  between  the black D3 brane  with and without D-instantons comes from the 
probe D-brane dynamics, which should be 
calculated in the string frame  where dilaton factor is manifest as an over all factor in the metric.  
It turns out that the Chern-Simons term can not cancel the effect of the dilaton 
unlike the claim of reference \cite{Ghoroku}, which is the origin of the chiral symmetry breaking in our paper.  On the other hand, in  \cite{Ghoroku}, the flat embedding is prefered for some reason and 
it was attributed to the cancellation of the dilaton effect  by the CS term. As we have shown in the appendix A, this is not the case and brane embedding can not be flat, which in turn induces the chiral symmetry breaking. We attribute such effect to the net repulsive force acting on probe D7 created by the D-instanton charges. Such breaking of the chiral symmetry is consistent with the gauge theory  where zero mode of the fermions in the D-instanton  background requests chiral symmetry breaking. 

For the treatment of chemical potential, we follow the method first suggested in \cite{KSZ,HT}  and developed in \cite{NSSY,KMMMT}. 
There are  two types of probe brane embedding:   
one describes quark phase, where the flavor brane 
touches   the black hole horizon and a baryon vertex is not allowed. 
Usually this embedding is called {\it black hole embedding}. 
If there are  strings  connecting the horizon and the D7, the latter is deformed 
 to a spiky brane to touch the horizon. 
 The other embedding is the one where D7 brane never touches the horizon. 
 On such brane, strings can be attached only if 
  a baryon vertex (BV) is allowed and present 
  so that the strings connect  the BV to D7.
   We find that there is a phase transition between two embeddings.
 We also find that there is a transition between the two black hole embeddings. 
The  binding energy of the baryon and the melting temperature in this background was studied in \cite{Ghoroku2,zhou-sin} by considering one baryon vertex connected to the boundary at infinity by $N_c$ strings.  We treat the baryonic medium using the method developed in \cite{Seo:2008qc} where each compact baryon vertex is joined 
 with a D7 probe brane through a funnel and  such configuration is smeared along the D3 direction.

One final comment here is about the nomenclature: usually instanton gas  is for weakly interacting far-separated instanton and liquid is for dense and non-trivial interaction. Here we are using the words  "instanton liquid"  since homogenous distribution would not be consistent with the well separated gas configuration. We do not focus on the interaction strength at all.

 The rest of the paper is planned as follows. 
 In section 2,   the background geometry is reviewed and  various embedding configuration was studied
 with zero  density and chiral condensations are calculated. 
  In section 3, we study how the embedding geometry changes as we vary 3 parameters: temperature, baryon/quark density and the  gluon condensation. We find that there can be two phases within quark phase: one chiral symmetry broken and the other chiral symmetry restored.
  In section 4, by calculating the free energy and chemical potential as function of density 
  we study the phase transition between the baryon and quark phases as well as the 
  phase transition between the two quark phases. We work out the complete  phase diagram both in canonical 
 and grand canonical ensemble. 
 In section 6, we give a  conclusion and future directions. 
 In the appendix show that the Chern-Simons term can not cancel the effect of the dilaton   unlike the reference \cite{Ghoroku}.

\section{Background geometry and D-brane setup}
Here we will briefly review  the background geometry and the probe brane setup within it. 
The geometry  is the one that is a finite temperature extension  of D3/D-instanton background with Euclidean signature given by\cite{Ghoroku}.
The background has  a five-form field strength and a  axion field which couples to D3 and D-instanton, respectively. The ten dimensional supergravity action in Einstein frame  is given by \cite{Gibbons:1995vg, Kehagias:1999iy}
\be\label{10daction}
S=\frac{1}{\kappa} \int d^{10}x \sqrt{g}\left( R-\frac{1}{2} \left(\partial \Phi \right)^2 +\frac{1}{2}e^{2\Phi} \left(\partial \chi \right)^2 -\frac{1}{6} F_{(5)}^2\right), 
\ee
where $\Phi$ and $\chi$ denote the
dilaton and  the axion respectively.
If we set $\chi = -e^{-\Phi} + \chi_0$,  the dilaton term cancel the axion term in (\ref{10daction}). Then the action becomes that of the metric and five-form.  The solution in string frame can be written as
\bea\label{10dmetric}
ds_{10}^2 &=& e^{\Phi/2}\left[ \frac{r^2}{R^2} \left(f(r)^2 dt^2 + d\vec{x}^2 \right)
+\frac{1}{f(r)^2} \frac{R^2}{r^2} dr^2 +R^3 d\O_{5}^2 \right],\cr &&
e^{\Phi}=1+\frac{q}{r_T^4}\log\frac{1}{f(r)^2},~~~~~~\chi = -e^{-\Phi} +\chi_0,\cr
&&~~~~~~~~~~~~f(r) =\sqrt{1-\left(\frac{r_T}{r}\right)^4},
\eea
where $R^4 =4\pi g_s  N_c {\alpha'}^2$. The constant $q$ denotes the number of D-instanton. From the AdS/CFT dictionary, it also represents the vacuum expectation value (vev) of gluon condensation.  The dilaton factor diverges at the black hole horizon. However, it does not 
give any effect on the thermodynamics of the {\it bulk  theory} 
since the latter should be calculated in the Einstein frame where its onshell action is the same as that of the black D3 brane solution. 
  The geometry has a regular event horizon and Hawking temperature given by $T=r_T/\pi R^2$. \par

Quark-antiquark potential is derived from the expectation value of a Wilson  loop  using  U-shaped fundamental string configuration \cite{Ghoroku}.  For a given temperature $T$, U-shaped  string touches the black hole horizon and splits into two straight strings  at  certain separation $L_*(T)$. This critical distance $L_*$ increases as temperature decreases as in the usual black hole geometry.  
The geometry corresponds to deconfined phase from the gluon point of view. 
In usual black D3 brane geometry, $q\bar{q}$ potential is Coulomb like for small separation 
 and flat when separation is larger than $L_*$, where  it is completely screened.
In this background, the potential is Coulomb like  at  short distances but  linearly grows  
until U-shaped string touches the horizon. At zero temperature the potential is linear up to indefinitely large distance. 
 This is a similarity  to the real QCD. 
 But the gluon is deconfined which is different from the real QCD. 
We call such phenomena as `quasi-confinment'. Such quasi-confining property comes from the presence of  $q$, which is the value of gluon condensation. It is the D-instanton number and it should be closely related to the chiral condensation as well as the gluon condensation. 
 \vskip .5cm
 
Introducing a dimensionless coordinate $\xi$  by $\frac{d\xi^2}{\xi^2}=\frac{dr^2}{r^2f^2(r)}$,
the background geometry  can be rewritten as 
\be\label{10dmetric2}
ds^2 =
e^{\Phi/2}\left[\frac{r^2}{R^2}\left(f(r)^2 dt^2 +d\vec{x}^2\right)+\frac{R^2}{\xi^2}
\left(d\xi^2 +\xi^2 d\O_5^2 \right)\right], 
\ee where $r$ and $\xi$ are related by 
\be\label{fxi}
\left(\frac{r}{r_{T}}\right)^2 = \frac{1}{2}\left(\frac{\xi^{2}}{\xi_T^2}+\frac{\xi_T^2}{\xi^{2}}\right), \;\; {\rm
and } \;\; f = \left(\frac{1-\xi_T^4/\xi^{4}}{1+\xi_T^4/\xi^{4}}\right) \equiv \frac{\omega_{-}}{\omega_{+}}, ~~~\o_{\pm}\equiv1\pm\frac{\xi_T^4}{\xi^{4}}.
\ee
\par
To describe the embedding of the 
probe D7 brane, 
we decompose $\mathbb{R}^6$ part in (\ref{10dmetric2}) into $\mathbb{R}^4 \times \mathbb{R}^2$,
\be
ds^2=e^{\Phi/2}\left[\frac{r^2}{R^2}\left(f(r)^2 dt^2 +d\vec{x}^2\right)+\frac{R^2}{\xi^2}
\left(d\rho^2 +\rho^2 \Omega_3^2+dy^2 +y^2 d\phi^2 \right)\right].
\ee
D7 brane spans $(t, \vec{x}, \rho)$ direction and wraps $S^3$, and is perpendicular to $y$ and $\phi$ direction. We can set $\phi=0$ using  the $SO(2)$ symmetry in $x^8,x^9$ plane. Then induced metric on D7 brane becomes
\be\label{inducedmetric}
ds_{D7}^2=e^{\Phi/2}\left[\frac{r^2}{R^2}\left(f(r)^2 dt^2 +d\vec{x}^2\right)+\frac{R^2}{\xi^2}
\left((1+y'^2)d\rho^2 +\rho^2 \Omega_3^2 \right)\right],
\ee
where $y'$ denotes to $\partial y(\rho)/\partial \rho$ and $\xi^2 =\rho^2 +y^2$.\par
The general action of probe D7 brane can be written as
sum of DBI action of D7 brane and the  Chern-Simons term.  The embedding dynamics of the D7 brane  gives the dependence of the chiral condensation on control parameters like temperature, density, quark mass etc.  as we will describe below.

\section{ Chiral condensation at zero density}
Here we study D7 brane embeddings at zero density but 
we first consider the zero temperature and zero density case 
to study $q$ dependence of the chiral condensation  and then 
consider the finite temperature to study 
chiral phase transition temperature as a function of gluon condensation. 

\subsection{Zero temperature limit}
In this limit, 10 dimensional geometry (\ref{10dmetric}) becomes near horizon geometry of D3/D-instanton system which preserve half of supersymmetry \cite{Liu:1999fc},
 \bea\label{10dmetric3}
ds^2 &=&
e^{\Phi/2}\left[\frac{r^2}{R^2}\left( dt^2 +d\vec{x}^2\right)+\frac{R^2}{r^2}
\left(dr^2 +r^2 d\O_5^2 \right)\right], \cr\cr
e^{\Phi}&=& 1+\frac{q}{r^4},~~~~~\chi =-e^{-\Phi} +\xi_{\infty}.
\eea
Induced metric on D7 brane can be written as
\be
ds_{D7}^2=e^{\Phi/2}\left[\frac{r^2}{R^2}\left(dt^2 +d\vec{x}^2\right)+\frac{R^2}{r^2}
\left((1+y'^2)d\rho^2 +\rho^2 d\Omega_3^2 \right)\right],
\ee
where $r^2 =\rho^2 +y^2$. The presence of D-instanton implies that of  axion. The axion can couple to the D7 brane world volume through the Chern-Simons term.  Therefore D7 brane action can be written as  
\bea\label{zeroTdvi}
S_{D7} &=& S_{DBI} +S_{CS} \cr\cr
&=&  -\mu_7 \int d\sigma^8 e^{-\Phi} \sqrt{-{\rm det}(g+2\pi \alpha' F)} +\mu_7\int d^8\sigma\frac{1}{8!} C_{(8)i_1 \cdots i_8},
\eea
where $\sigma$ is world volume coordinates of D7 brane, $g$ is induced metric on D7 brane and $C_{(8)}$ is the Hodge dual 8-form gauge potential of axion field which is 0 form. If we fix D7 brane position along the fixed $\phi$ direction, the Chern-Simons term becomes locally total derivative
as we proved in appendix  \ref{CS}. Therefore it can not affect  the equation of motion hence we can ignore the Chern-Simons  term. 
This observation makes the key difference of our D7 brane embedding to the one described in \cite{Ghoroku} and it will be the origin of the chiral symmetry breaking. 

Now, the DBI action for D7 brane can be written as
\be
S_{D7} = -\tau_7 \int dt d\rho  \,\,e^{\Phi} \rho^3 \sqrt{1+y'^2},
\ee
with  $\tau_7 =\mu_7 V_3 \Omega_3$ and the equation of motion is
\be\label{emo0}
\frac{d}{d\r}\left(\frac{e^{\Phi} \r^2 y'}{\sqrt{1+y'^2}}\right) +\frac{q\cdot y \r^3 e^{\Phi}\sqrt{1+y'^2}}{r^6} =0.
\ee
For a generic value of $q$,   analytic solution is not available. So we look for  numerical one.  For $q=0$,  the trivial embedding $y={\rm constant}$   is a solution  which is consistent with the result   
in \cite{Kruczenski:2003be}. \par 

\begin{figure} [ht!]
\begin{center}
\subfigure[]{\includegraphics[angle=0,width=0.4\textwidth]{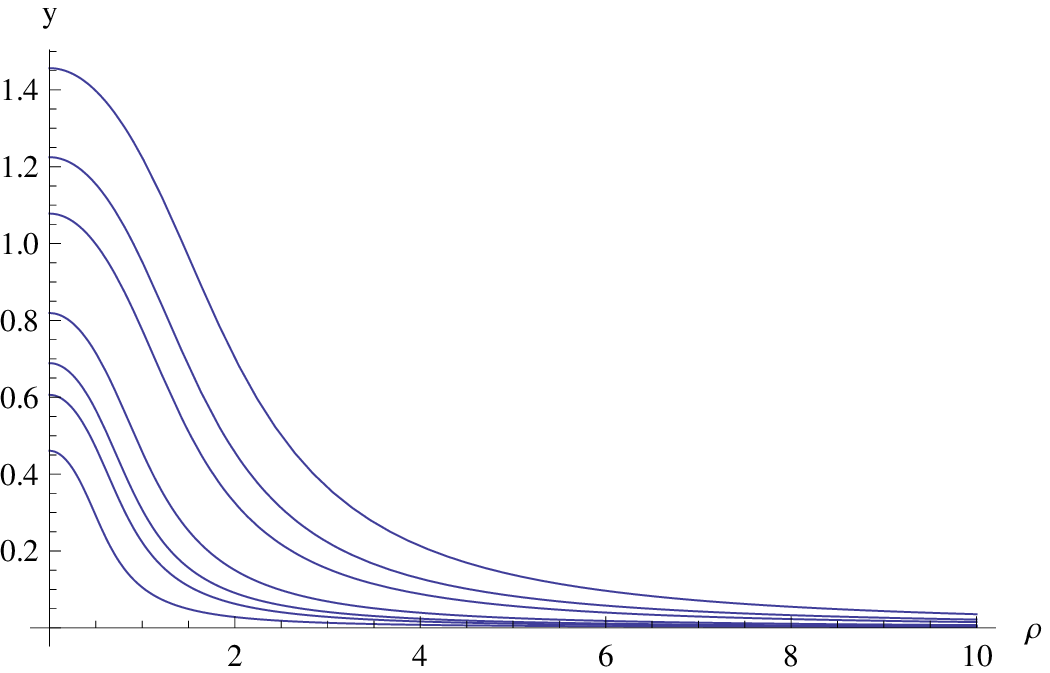}} 
\hspace{1cm} 
\subfigure[]{\includegraphics[angle=0,width=0.4\textwidth]{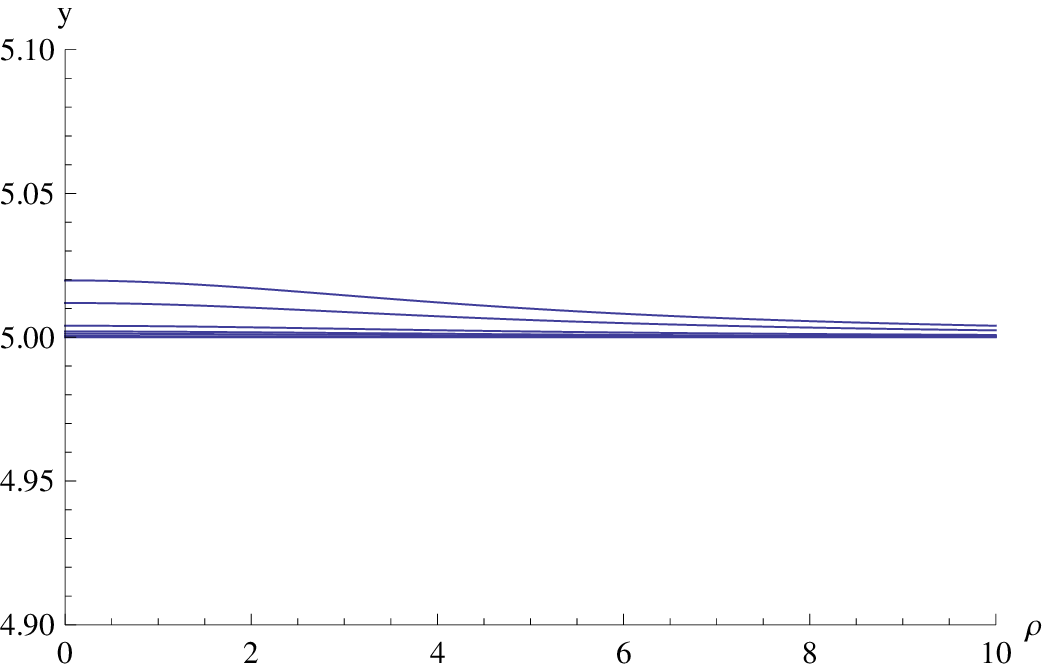}} \caption{(a) D7 brane embeddings for $m_q =0$ with $q=0,~0.1,~0.3,~0.5,~1,~3,~5,~10$ from bottom to top. (b) D7 brane embeddings for $m_q =5$ with $q=0,~0.1,~0.3,~0.5,~1,~3,~5,~10$ from bottom to top .
\label{fig:zeroTQd7}}
\end{center}
\end{figure}

\begin{figure} [ht!]
\begin{center}
\subfigure[]{\includegraphics[angle=0, width=0.4\textwidth]{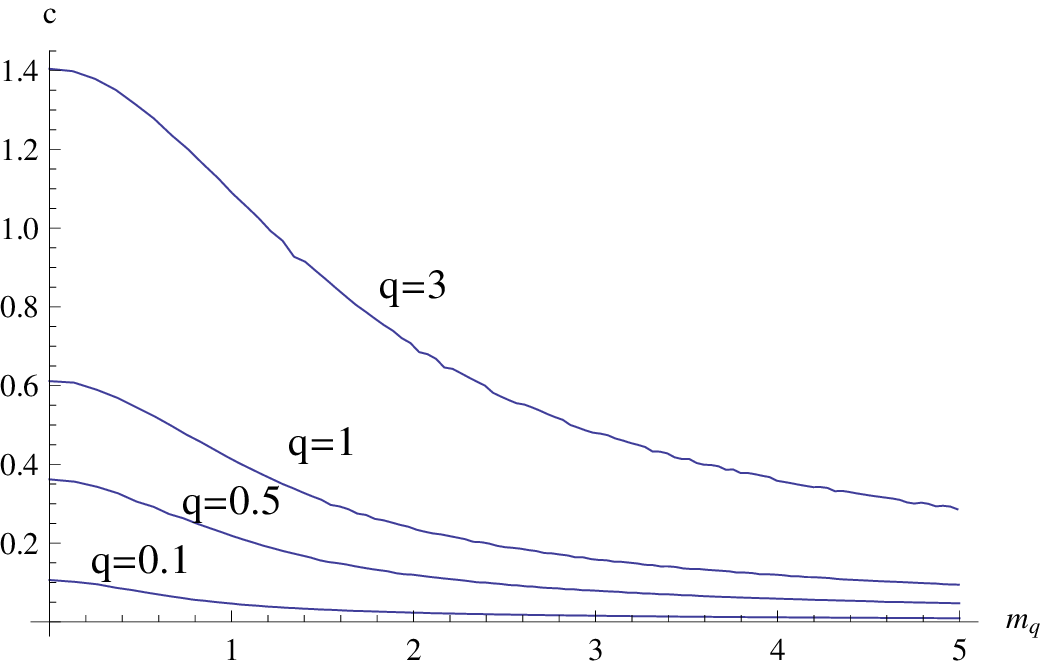}}
\hspace{1cm}
\subfigure[]{\includegraphics[angle=0, width=0.4\textwidth]{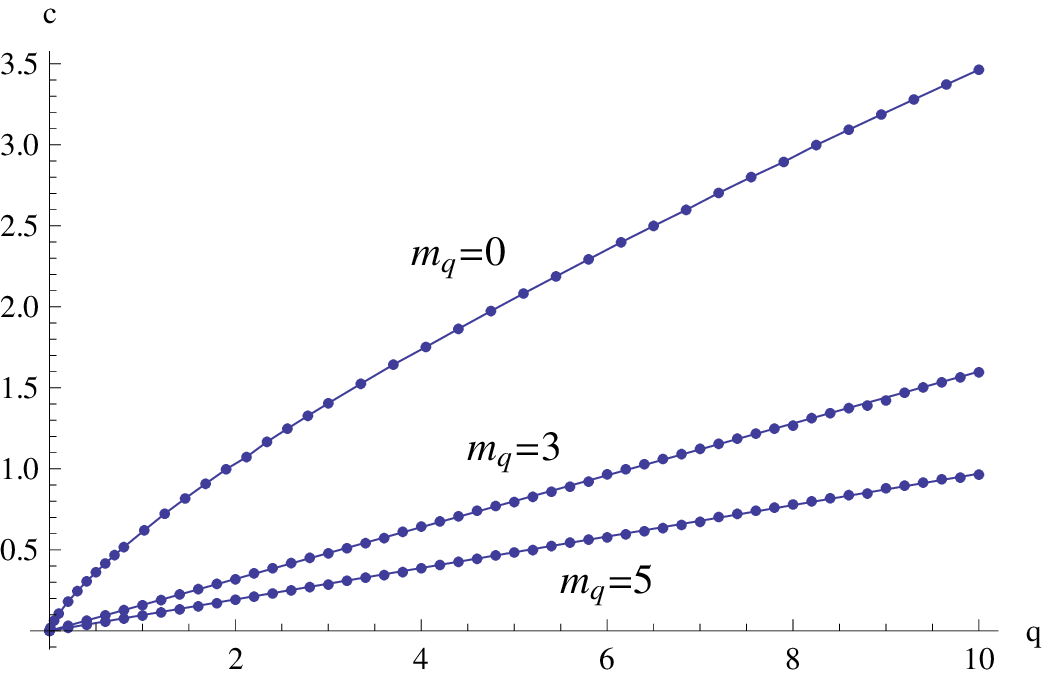}}
\caption{(a) $m_q$ dependence of chiral condensation  with fixed $q$. (b) $q$ dependence of chiral condensation with fixed $m_q$.
\label{fig:zeroTQ_cond}}
\end{center}
\end{figure}

If we turn on $q$, the solution deforms. The bare quark mass $m_q$ and chiral condensation $c$  
are encoded in the  asymptotic form of embedding function $y(\rho)$ \cite{Evans,Kruczenski}: 
\be\label{yrho}
y(\rho) = m_q + \frac{c}{\rho^2} +\cdots.
\ee
The embeddings with fixed $m_q$ are drawn in Figure \ref{fig:zeroTQd7}. As $q$ increases, probe D7 brane  in the centeral region bends  upward more and more so that we have non-vanishing chiral condensation that is an increasing function of $q$. The value $q$ corresponds to the gluon condensation $<{\rm Tr} F^2>$ in boundary theory and hence plays the role of parameter of scale symmetry breaking.  The $m_q$ dependence of chiral condensation is drawn in Figure \ref{fig:zeroTQ_cond}(a). With non-zero value of $q$, the value of chiral condensation goes to finite value in $(m_q \rightarrow 0)$. Therefore, the chiral symmetry is broken. 
 The $q$ dependences of the chiral condensation for given $m_q$ are drawn in Figure \ref{fig:zeroTQ_cond}(b). The  condensation is  increasing function of $q$ but   decreasing one of $m_q$. 
 At large quark mass, chiral condensation is linearly in $q$.  It is consistent with   the expectation from field theory \cite{Shifman,Kruczenski,erdmenger},
\be
<\bar{\psi}\psi> =\frac{\alpha_s N_f}{12 \pi m_q} <{\rm Tr} F^2>.
\ee

\subsection{Finite temperature without density}
In this section, we will study D7 brane embedding in black hole geometry without chemical potential. We use the black hole metric (\ref{10dmetric}) and induce metric on D7 brane with (\ref{inducedmetric}). Then DBI action for D7 brane becomes
\bea
S_{D7} &\equiv& \t_7 \int dt d\rho V(\rho,y) \sqrt{1+y'^2},
\eea
where $\tau_7 = \xi_T^4 \mu_7 V_3 \Omega_3$ and $V= e^{\Phi}\rho^3 \omega_+ \cdot\omega_-  $. 
The dilaton factor seems to diverge at the black hole horizon. However, the dilaton factor $e^\Phi$ comes with  $\omega_-$  which contains the  zero  $\sim (r-r_T)$ at the horizon. This zero  is enough to kill the the logarithmic divergence of the dilaton. In other words,  the dilaton's  log singularity does not change the qualitative behavior of the brane dynamics near the horizon. 

\begin{figure} [ht!]
\begin{center}
\subfigure[]{\includegraphics[angle=0, width=0.4\textwidth]{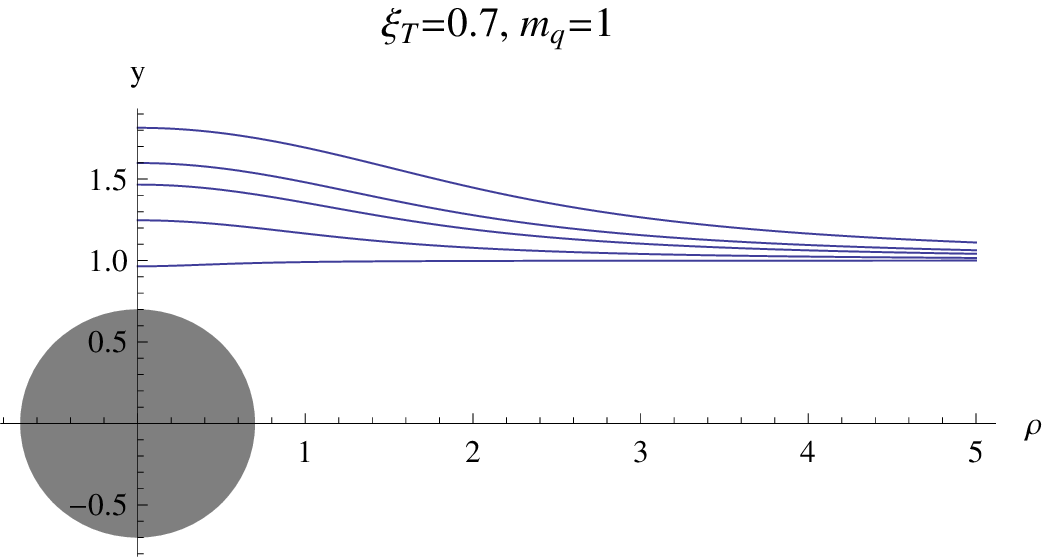}}
\hspace{1cm}
\subfigure[]{\includegraphics[angle=0,width=0.35\textwidth]{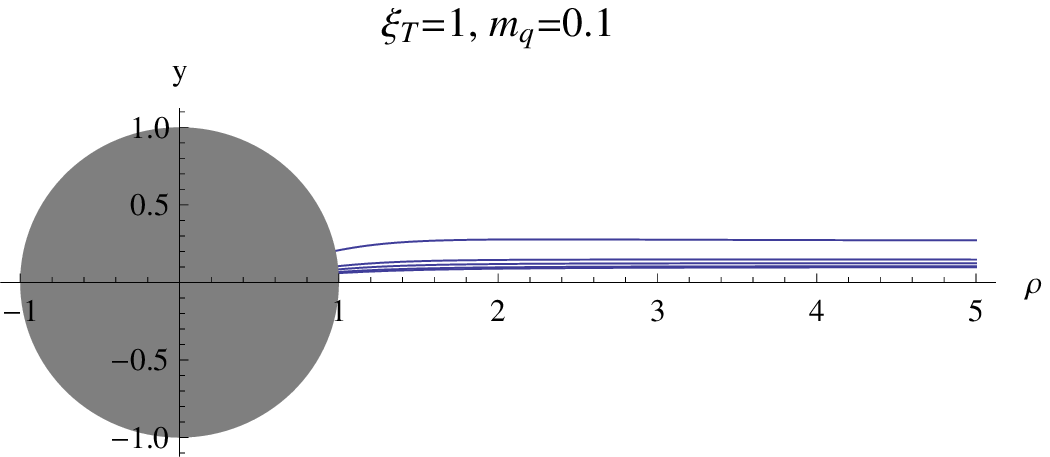}} 
 \caption{(a) $q$ dependence of Minkowski embedding for $m_q=1$ with $q=0,~1,~3,~5,~10$ from below. (b) $q$ dependence of black hole embedding for $m_q=0.1$ with $q=0,~1,~3,~5,~10$ from below.
\label{fig:zeroQ01}}
\end{center}
\end{figure}

 The equation of motion for the DBI action can be written following form
\be
\frac{y''}{1+y'^2} +\frac{\partial \log V}{\partial \rho} y' -\frac{\partial \log V}{\partial y}=0.
\ee
This equation of motion is highly non-linear.  We can get solution only  in numerical way  with  proper boundary condition (BC). In the presence of black hole horizon, embedding of probe D7 can be classified as   'Minkokwski embedding' and 'black hole embedding' \cite{Mateos:2006nu}. For the Minkowski embedding,  we impose BC:   $y(0)=y_0$ and $y'( 0) =0$. For the black hole embedding, BC is determined by regularity condition of equation of motion at the horizon:
\be
y( \rho_{min})=y_0,~~~~~y'( \rho_{min}) = \tan \th,
\ee
where $\theta$ is the angle between $\rho$ axis and probe brane position at the horizon.  In usual black hole geometry, gravitational attraction of black hole bends the probe brane downward.  However,  non-zero value of $q$ gives net  `repulsive' force on probe D7. Therefore, D7 bends upward. 
$q$  dependences of Minkowski embedding and black hole embedding are drawn in Figure \ref{fig:zeroQ01}. For fixed quark mass and temperature, the bigger is $q$, the more   pushed up is the D7 brane.  
Such effect is common both in Minkowski  and  black hole embedding. 
 
\begin{figure} [ht!]
\begin{center}
\includegraphics[angle=0, width=0.4\textwidth]{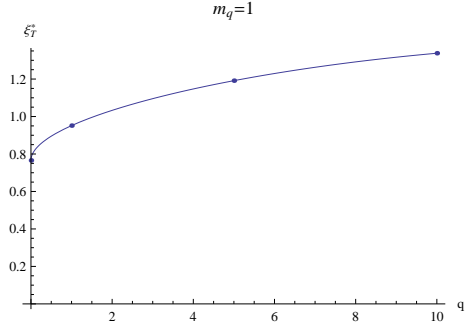}
\caption{$q$ dependence of phase transition temperature.
\label{fig:PT_zeroQ2}}
\end{center}
\end{figure}

We expect a phase transition between Minkowski  and black hole embeddings as we increase temperature.
Namely, Minkowski embedding in low temperature will change into black hole embedding  in high temperature. 
In the presence of $q$, D7 feels repulsive force, therefore we expect that as $q$ increase, phase transition temperature goes up.
One may expect that the phase transition might be smoother  compared with  the case of no instanton charge. 
One can extract  $q$ dependence of phase transition temperature $\xi_T^*$ from the free energy.
The case for   $m_q =1$ is drawn in  Figure \ref{fig:PT_zeroQ2}.

\begin{figure} [ht!]
\begin{center}
\subfigure[]{\includegraphics[angle=0,width=0.3\textwidth]{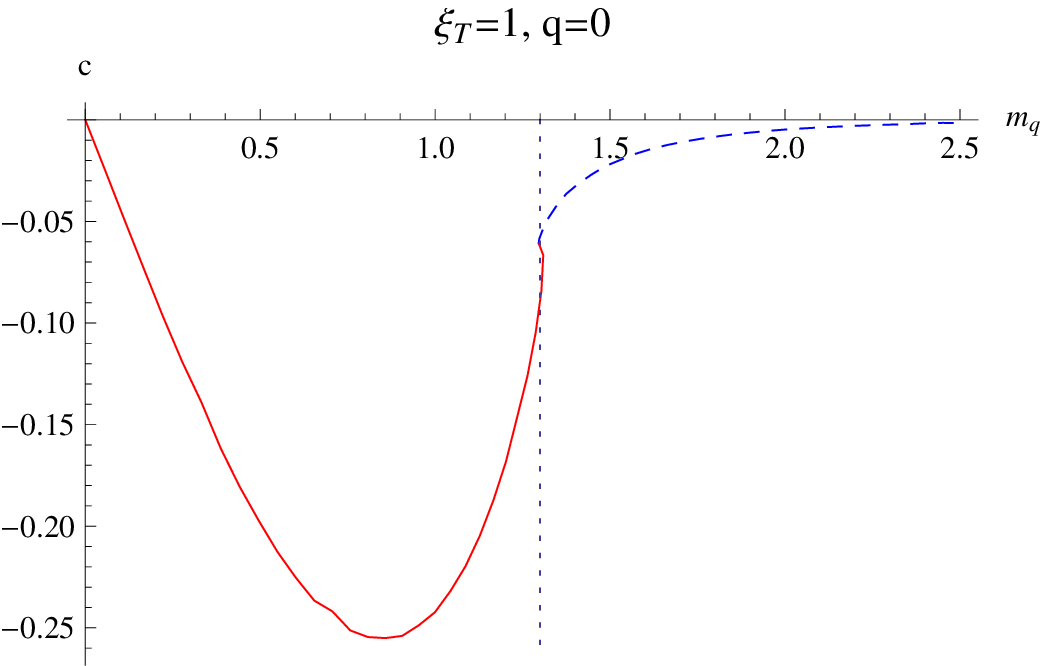}} 
\subfigure[]{\includegraphics[angle=0,width=0.3\textwidth]{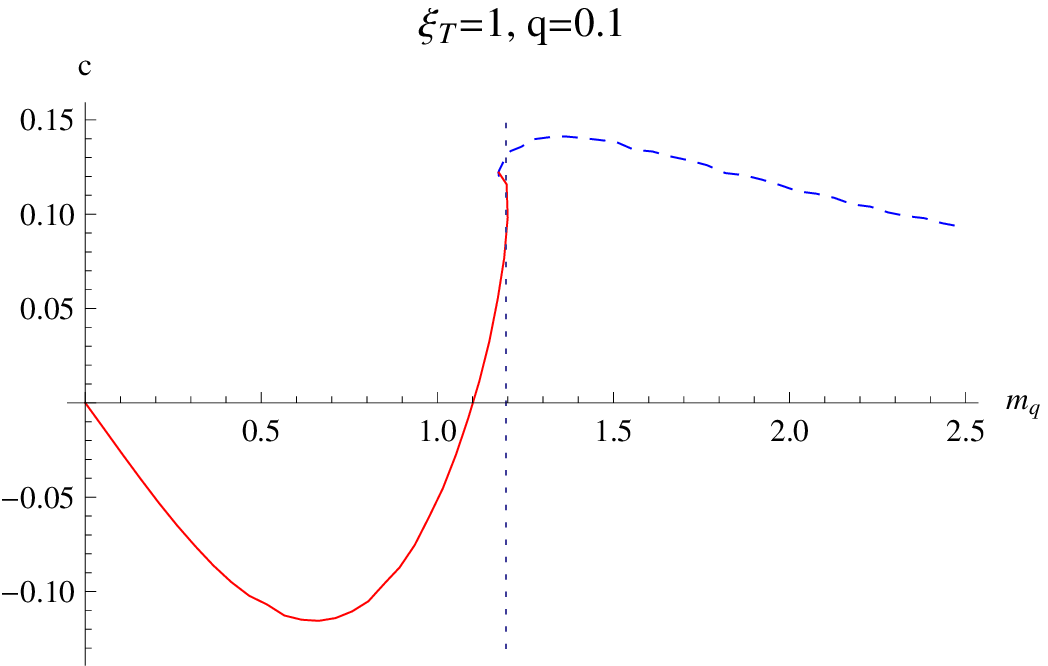}} 
\subfigure[]{\includegraphics[angle=0,width=0.3\textwidth]{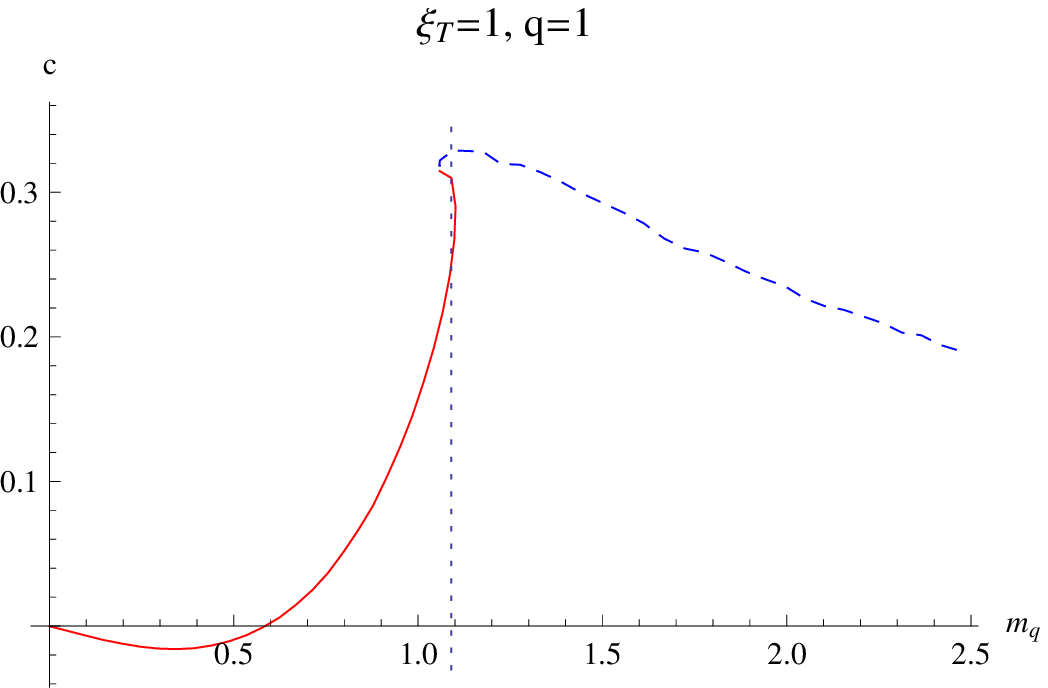}} 
\subfigure[]{\includegraphics[angle=0,width=0.3\textwidth]{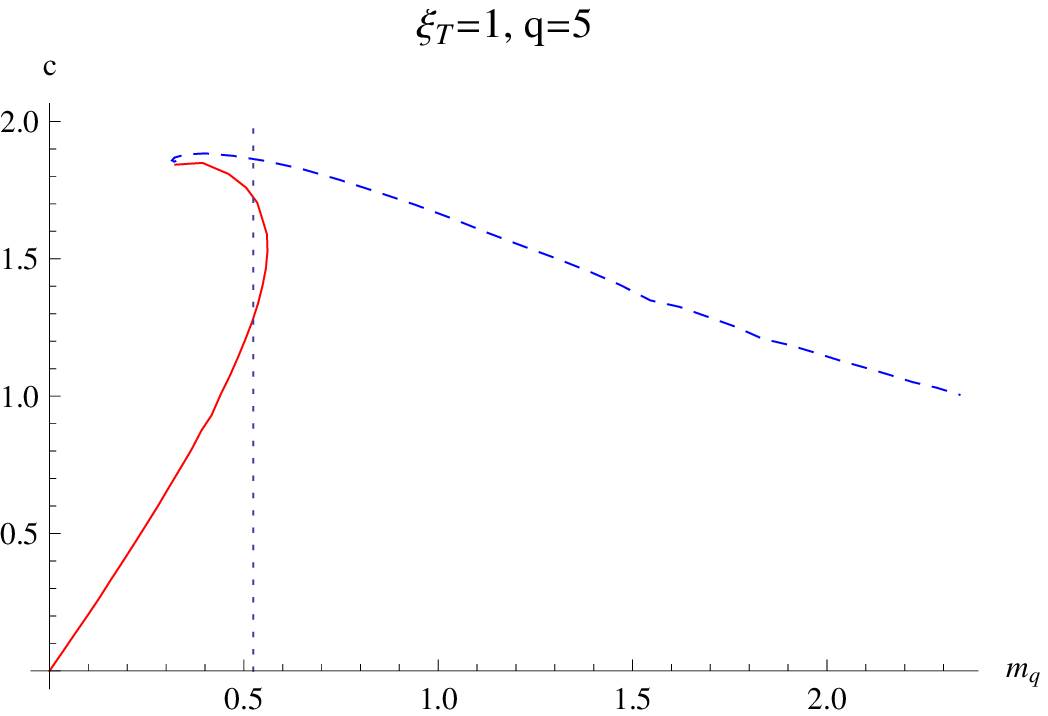}} 
\subfigure[]{\includegraphics[angle=0,width=0.3\textwidth]{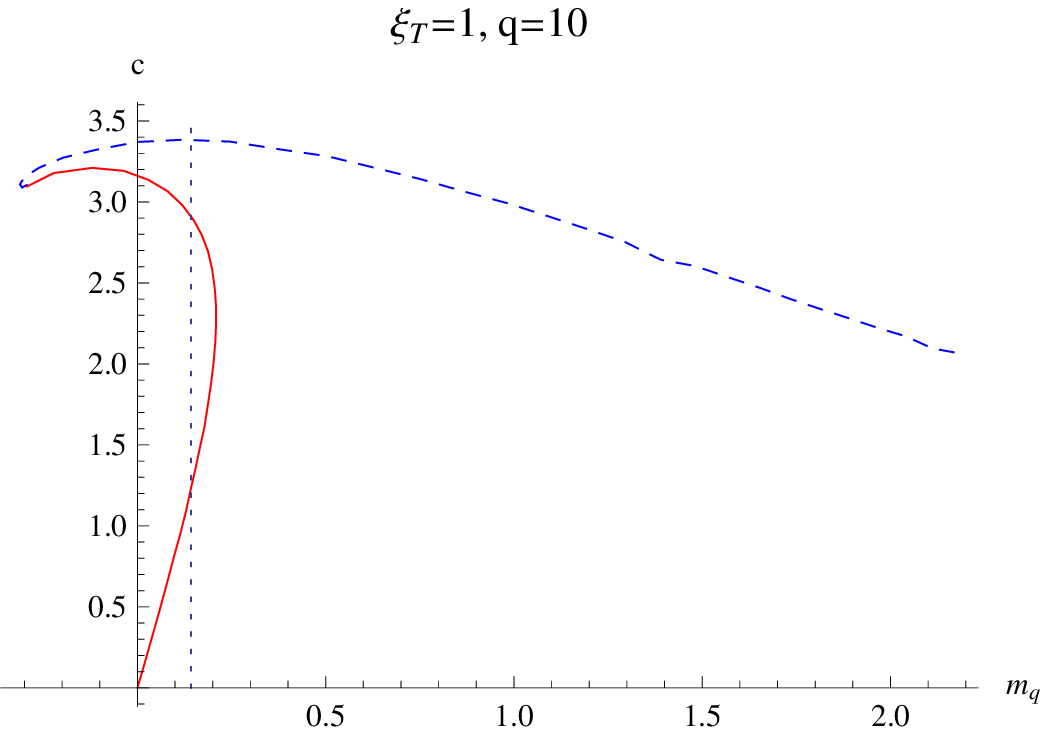}} 
\subfigure[]{\includegraphics[angle=0,width=0.3\textwidth]{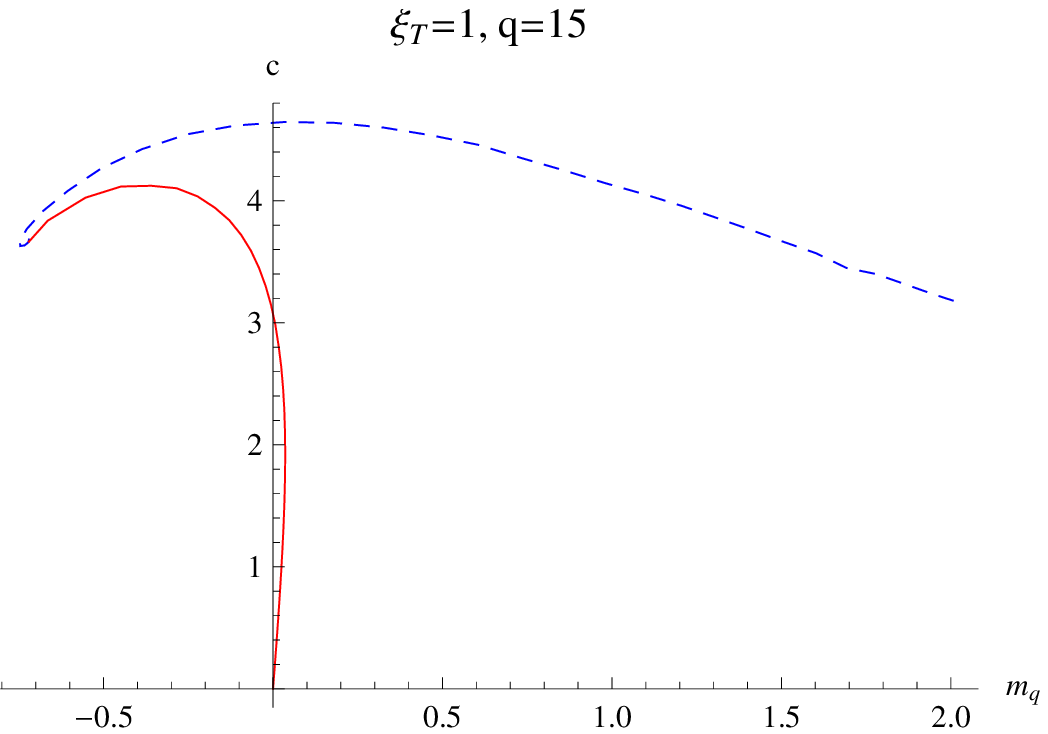}} 
\caption{$m_q$ dependence of chiral condensation $c$  for $\xi_T =1$. Blue dashed line denotes to chiral condensation for Minkowski embedding and red line denotes to black hole embedding. Dotted line indicates phase transition between two embeddings.
\label{fig:cmq}}
\end{center}
\end{figure}

Now, we will discuss about  chiral condensation for this system:  
it was defined as the slope of D7 brane at asymptotic region. See (\ref{yrho}). We first fix the temperature  and calculate  it  with different $m_q$'s. 
For $q=0$, D7 brane bends down  and its slope at the asymptotic region is  positive and 
 therefore the condensation value is  negative. See Figure \ref{fig:cmq}(a). 
If we turn on $q$ and increase it slowly, the brane  bends  up  relatively  very small value of $q$
since the pushing-up effect of $q$ is very effective. 
Therefore the  sign of the condensation  flips  at small value of $q$. 
Also the value of $m_q$ where phase transition occur decreases as $q$ increases. 

\begin{figure} [ht!]
\begin{center}
\subfigure[]{\includegraphics[angle=0,width=0.3\textwidth]{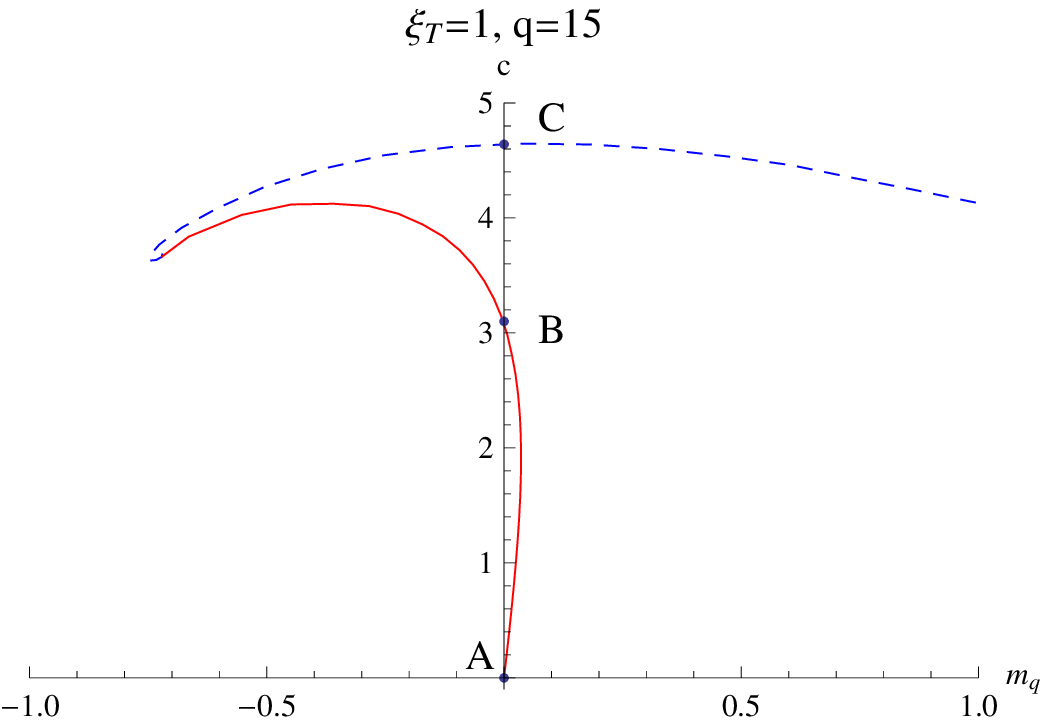}} 
\subfigure[]{\includegraphics[angle=0,width=0.3\textwidth]{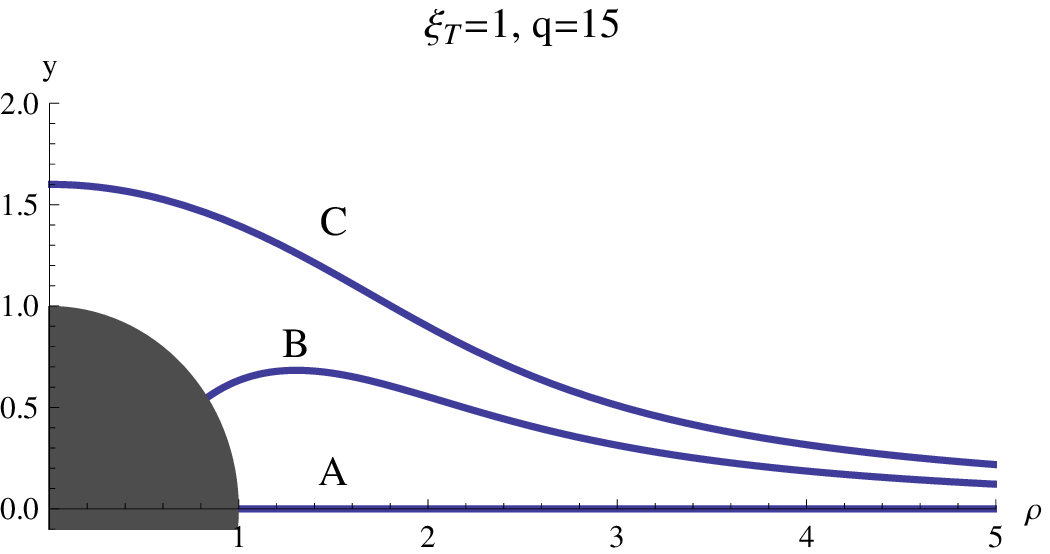}}  
\subfigure[]{\includegraphics[angle=0,width=0.3\textwidth]{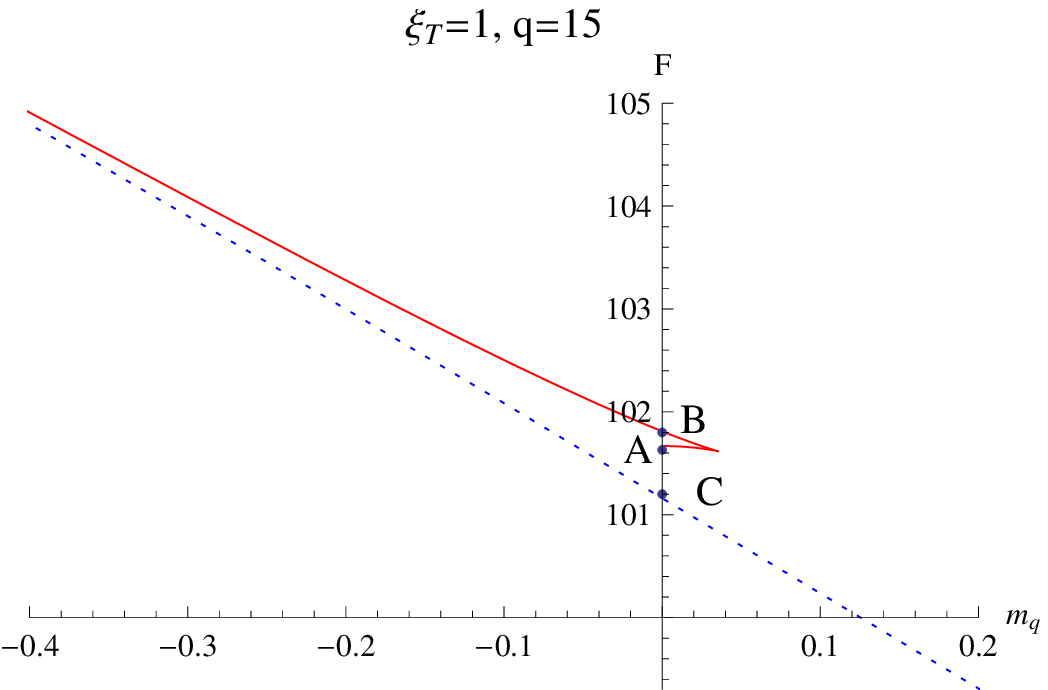}} 
\caption{(a) Figure 5(f). A, B and C denote three solution which give zero quark mass. (b) Probe brane embeddings for each case. (c) Free energy as a function of quark mass. C(Minkowski) embedding has minimum energy.
\label{fig:mq0}}
\end{center}
\end{figure}

As we decrease $m_q$ with fixed $q$, 
there is phase transition from Minkowski embedding to black hole embedding at certain value of $m_q$. 
See Figure \ref{fig:cmq}.  
In Figure \ref{fig:cmq} (e,f) there is   range where $m_q$ becomes negative. 
If we calculate free energy, there is phase transition before $m_q$ becomes negative.
Moreover, in Figure \ref{fig:cmq}(f), free energy of Minkowski embedding is always smaller than free energy of black hole embedding. It means that in $m_q \rightarrow 0$ limit, the value of chiral  condensation does not vanish, see Figure \ref{fig:mq0}. This is a spontaneous breaking of a U(1) symmetry which is analogue of the chiral symmetry  discussed \cite{Evans,Kruczenski}. 
Although the relevant symmetry is rotation in $x^8,x^9$ plane which is not a true chiral symmetry, 
one  can develop  the Gellman-Oakes-Renner (GOR)  relation \cite{Evans, Kruczenski}, which is the purpose 
of having the chiral symmetry.  Therefore from now on we 
call this as chiral symmetry breaking and we call  $c$  as the chiral condensation.

\begin{figure} [ht!]
\begin{center}
\includegraphics[angle=0, width=0.4\textwidth]{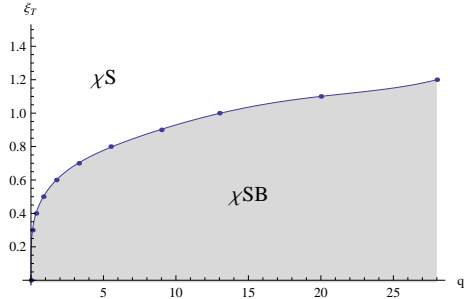}
\caption{$q$ dependence of chiral symmetry restoring temperature. $\chi$SB denotes  chiral symmetry breaking phase and $\chi$S to chiral symmetry restored phase
\label{fig:chiralB}}
\end{center}
\end{figure}

For a given temperature, the  chiral symmetry is broken, if $q$ is  large enough.
The $q$ dependence of chiral symmetry restoring temperature is drawn in Figure \ref{fig:chiralB}. When $q \rightarrow 0$, chiral symmetry restoring temperature goes to zero. 

\section{Chiral symmetry at Finite temperature and finite baryon density}
In this section, we   discuss  D7 brane embedding with finite density using induced metric (\ref{inducedmetric}). Adding density corresponds to turning on $U(1)$ gauge field $A_t (\rho)$ on D7 brane. The DBI action of D7 brane can be written 
\bea\label{dbid7}
S_{D7} &=& -\tau_7 \int dt d\rho \rho^3
e^{\Phi/2}\omega_+^{3/2}\sqrt{e^{\Phi/2} \frac{\omega_-^2}{\omega_+}\left(1+\dot{y}^2\right) -\tilde{F}^2} := \int dt d\r {\cal L}_{D7},
\eea
where
\be
\tau_7 =\mu_7 V_4 \Omega_3,~~~\tilde{F}=2\pi\alpha' F_{t\rho}
\ee
and dot denotes the derivative with respect to $\rho$. We used $A_{\rho}=0$ gauge. 
For a fixed charge dynamics, we need  Legendre transformation of Lagrangian, which we call 
 `Hamiltonian': 
\bea\label{HD7}
{\cal H}_{D7} = \tilde{F}\frac{\partial {\cal L}_{D7}}{\partial {\tilde F}} - {\cal L}_{D7} 
= \tau_7  \sqrt{e^{\Phi}\frac{\omega_-^2}{\omega_+} (1+\dot{y}^2)}\sqrt{\hat{Q}^2 +\rho^6 e^{\Phi}\omega_+^3},
\eea
where $\hat{Q} =Q/(2\pi \alpha' \tau_7)$, $Q$ is the number of source charges. 
Notice that near the horizon, the last factor is dominated by the dilatonic term and the whole action is reduced to the case of $Q=0$. 
Therefore  the argument for the regularity near the horizon 
goes  exactly the way  of the previous section. 
We can get numerical solution for the equation of motion 
provided we have proper boundary conditions. The stringy objects  corresponding to the sources on D7 brane is the end points of fundamental strings. 

Unlike usual black hole background, our background permits the presence of  baryon vertex.
Therefore there are two way of attaching fundamental strings on D7 brane. 
One is  connecting  D7 to the black hole horizon and the other is  connecting it to the baryon vertices.
Two configurations give  different boundary conditions for the D7 brane dynamics.
We need to examine which configuration has lower free energy. 

\subsection{Quark phase}
One way to put point electric sources on D7 brane is to add fundamental strings such that one end of strings are on D7 and the other end on black hole horizon. It is equivalent to add freely moving quark in boundary theory because the fundamental strings can move freely on both D7 and horizon. We call it `quark' phase. Since the tension of D7 brane is always smaller than that of fundamental string\cite{KMMMT}, D7 brane pull down to the horizon.  \par
\begin{figure} [ht!]
\begin{center}
\subfigure[]{\includegraphics[angle=0, width=0.4\textwidth]{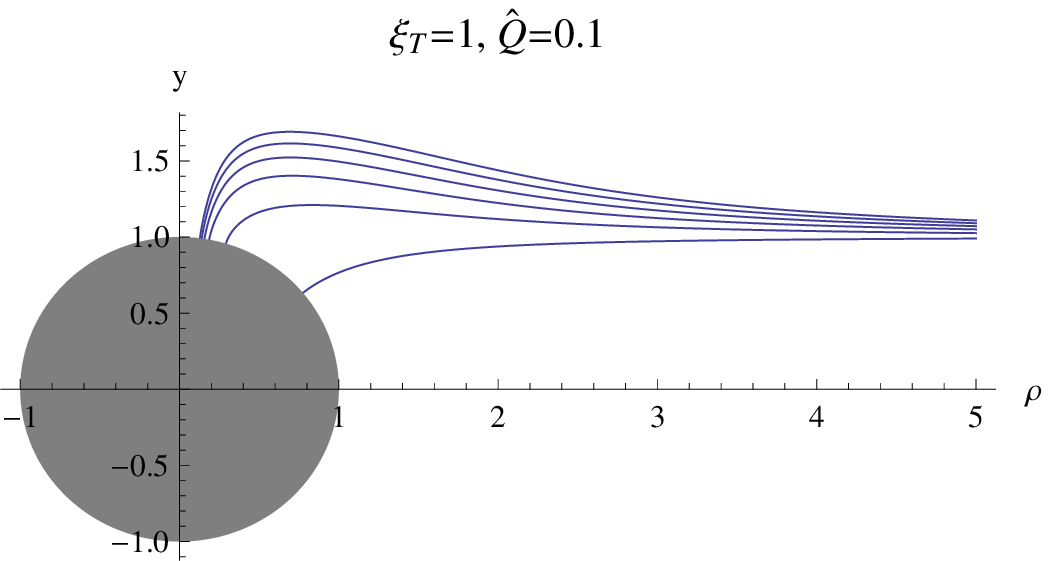}}
\hspace{1cm} 
\subfigure[]{\includegraphics[angle=0,width=0.4\textwidth]{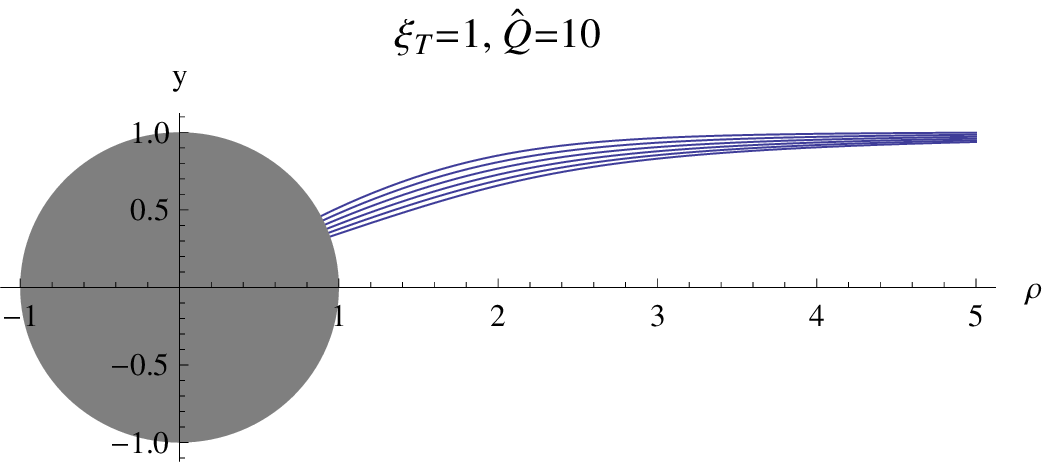}} 
\caption{D7 brane embeddings for $m_q =1$ and $\xi_T=1$ with $q=0,~2,~4,~6,~8,~10$ from below for (a) $\hat{Q} =0.1$, (b) $\hat{Q}=10$ \label{fig:BE1}}
\end{center}
\end{figure}

Regularity at the black hole horizon  requests
\be
\dot{y}(\rho_{min})=\tan\theta,
\ee
where $\theta$ is polar angle of the position  where D7 brane touch the horizon. As discussed in the previous section, the presence of $q$ gives repulsion on probe D7 brane, and it affects its embedding.  The $q$ dependence of   D7 brane embeddings are drawn in Figure \ref{fig:BE1}
for  two different density $Q$.

From the figure, we can see that as $q$ increases, the repulsion effect on D7  also increases in small density $\hat{Q}$. However, if $\hat{Q}$ is not small, the brane embedding is less sensitive to $q$ as shown in Figure \ref{fig:BE1}(b). 
This is because the charge $\hat{Q}$ introduces flux whose electric field energy increases the tension of the D7. 
That is, the stiffness due to the flux is dominating repulsion due to the $q$.\par

\begin{figure} [ht!]
\begin{center}
\subfigure[]{\includegraphics[angle=0, width=0.45\textwidth]{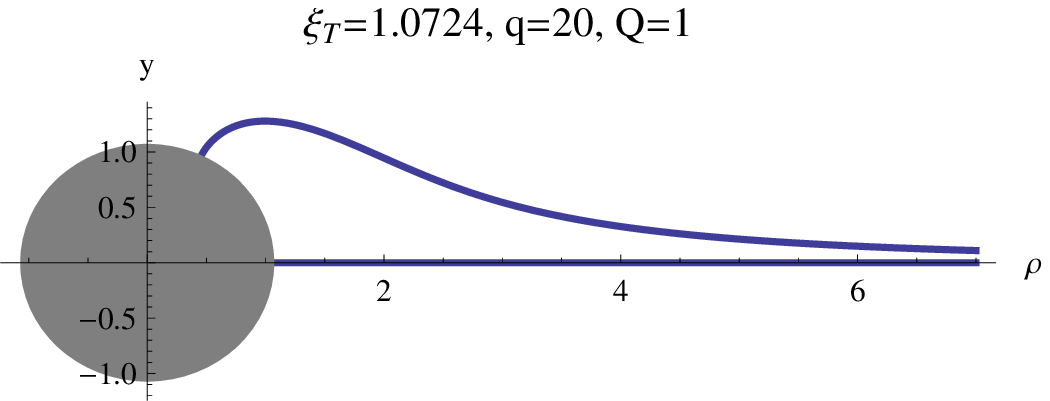}}
\hspace{1cm} 
\subfigure[]{\includegraphics[angle=0,width=0.45\textwidth]{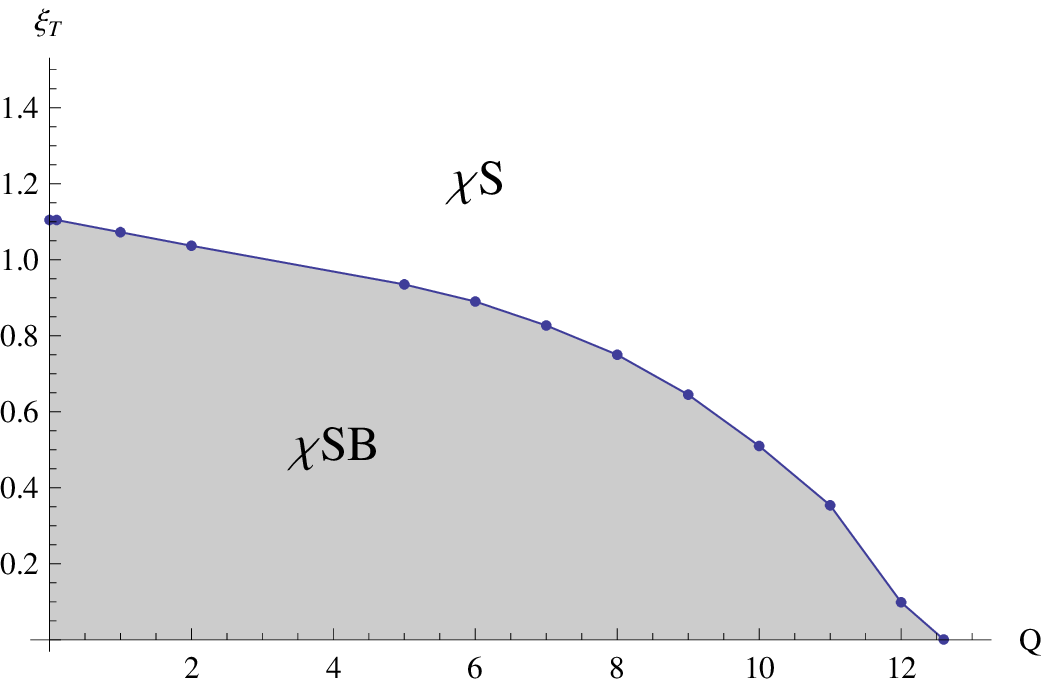}} 
\caption{(a) Two embedding solutions with same temperature, density and $q$.  (b) Phase boundary between chiral symmetry breaking and restored phases with $q=20$.  \label{fig:bhsol}}
\end{center}
\end{figure}

\begin{figure} [ht!]
\begin{center}
\includegraphics[angle=0, width=0.4\textwidth]{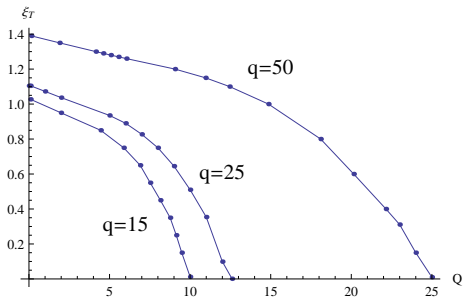}
\caption{phase diagram of chiral symmetry restoration within quark phase  for q=15, 25  and 50.
\label{fig:chiralbds}}
\end{center}
\end{figure}

In the absence of $q$, embedding solution which corresponds to $m_q=0$ is uniquely determined to be the flat embedding, $y(\rho)=0$ for which  value of chiral condensation is automatically zero. However, in the presence of $q$, there are two  different embeddings  for the given $m_q$. For $ m_q=0$, one is trivial with zero chiral condensation $c$ and the other has non-zero chiral condensation 
as we can see in Figure \ref{fig:bhsol}(a). The solution with non-zero $c$ exists in low temperature and small density region. In large density region, only $y=0$ is the solution.  Actually we found that there is a phase transition between the two embeddings  in certain temperature and density. The phase diagram is presented in Figure \ref{fig:bhsol}(b).  
 As $q$ increase  the phase boundary expands toward larger temperature and density region. See Figure \ref{fig:chiralbds}.
 This is natural since the chiral symmetry breaking is  caused by the effect of gluon condensation $q$, as we have seen in Figure 
 \ref{fig:zeroTQ_cond}\footnote{This phenomenon is very similar when we turn on the magnetic field on probe brane \cite{Evans:2010iy} in black D3 brane background. But in that case, the magnetic field cannot affect the background geometry and the interpretation of magnetic field also different.}.

\begin{figure} [ht!]
\begin{center}
\includegraphics[angle=0, width=0.4\textwidth]{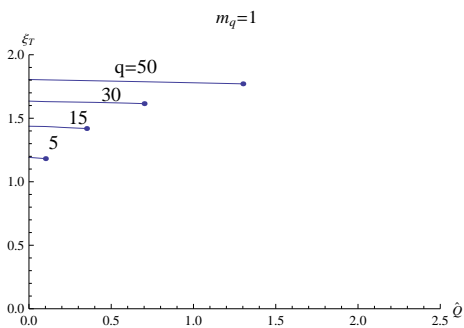}
\caption{$q$ dependence of phase transition between two quark phase for fixed $m_q =1$. These phase transition lines end with second order phase transition point.
\label{fig:btobs}}
\end{center}
\end{figure}

In Figure \ref{fig:BE1},  D7 embedding  for non-zero $m_q$ is drawn. In this case, behavior of D7 brane embedding is similar to black hole embedding with $q=0$.  Due to the non-zero value of $y(\infty)\sim m_q$, chiral symmetry is explicitly broken.   However, it is known that there is first order phase transition in black hole background between two quark phases in small density region\cite{NSSY,KMMMT}. This phase transition line finishes at  certain density and temperature. At this point, the order of phase transition is second. As $q$ increase, the phase phase boundary line  moves upward in (T,Q) plane. See Figure \ref{fig:btobs}.

 There is an issue about thermodynamical instability around first order phase transition point in black hole background \cite{NSSY,KMMMT}: 
chemical potential decreases as density increases, that is,  $\partial{\mu}/\partial{Q} <0$. But in the presence of $q$, this instability can be cured as we will discuss later.
 
\subsection{Baryon phase}
In this section, we study a baryon vertex in the D3/D-instanton background. Baryon vertex is originally proposed in \cite{Witten:baryon}. Near horizon geometry of black D3 brane is $AdS_5 \times S^5$. If we wrap spherical D5 brane on $S^5$, due to the Chern-Simons interaction between R-R five form field strength and D5 brane world volume, $U(1)$ gauge field is induced on D5 brane world volume and to cancel these fluxes we need to put $N_c$ fundamental strings on D5 brane. In asymptotic region, this object looks like bound state of $N_c$ fundamental string or quark. We call this object to `baryon vertex'. \par
As discussed in \cite{Seo:2008qc}, however,
in Schwartzschild type black hole background does not allow compact D5 brane as a solultion of equation of motion of DBI action. Therefore, the quark phase is only physical in finite temperature system. But, in the  D3/D-instanton background, compact D5-brane with $N_c$ fundamental strings can be formed
 even in the black hole background.\cite{Ghoroku2}. By connecting baryon vertex and probe D7 brane \cite{Seo:2008qc}, we can discuss thermodynamics of finite density and temperature system.\par

To study properties of baryon vertex, we rewrite the 10 dimensional metric (\ref{10dmetric2});
\be
ds^2 =
e^{\Phi/2}\left[\frac{r^2}{R^2}\left(f(r)^2 dt^2 +d\vec{x}^2\right)+R^2
\left(\frac{d\xi^2}{\xi^2} +d\th^2 +\sin^2\th d\O_4^2 \right)\right],
\ee
We take
$(t,\theta_{\a})$ as  world volume coordinates of a compact D5 brane, and turn on the $U(1)$
gauge field on it to have $F_{t\theta}\ne 0$.
 As the ansatz for the embedding of compact D5, we assume the $SO(5)$ symmetry so that position of D5 brane and
 gauge field depend only on $\theta$ i.e. $\xi=\xi(\theta)$, $A_{t}=A_t(\theta)$, where $\theta$  measure
 the polar angle of $S^5$ from the north pole. The induced metric on D5 brane is
\be
ds_{D5}^2 = e^{\Phi/2}\left[\frac{r^2}{R^2}f^2 dt^2 +
R^2\left(\frac{\xi'^2}{\xi^2}+1\right)d\th^2 +R^2 \sin^2\th d\O_4^2\right],
\ee
where
$\xi'=d\xi/d\th$. The DBI action for single D5 brane with $N_c$ fundamental string can be
written as
\bea\label{bary-d5}
S_{D5} &=& -\mu_5 \int e^{-\Phi} \sqrt{-{\rm det}(g+2\pi \alpha'
F)}+\mu_5 \int  A_{(1)}\wedge G_{(5)} \cr\cr 
&=& \t_5 \int dtd\theta \sin^4\theta e^{\Phi}
\left[-\sqrt{e^{\Phi} \frac{\o_-^2}{\o_+} (\xi^2 +\xi'^2)-\tilde{F}^2} +4 \tilde{A}_t
\right] \cr\cr 
&=& \int dt d\theta{\cal L}_{D5}, 
\eea 
where 
\be
\t_5 = \mu_5 \O_4 R^{4}r_T,~~~~~~
\tilde{F} = 2\pi \a' F_{t\theta},~~~~~~ \tilde{A}_t= 2\pi\a'A_t.
\ee

\begin{figure} [ht!]
\begin{center}
\subfigure[]{\includegraphics[angle=0, width=0.2\textwidth]{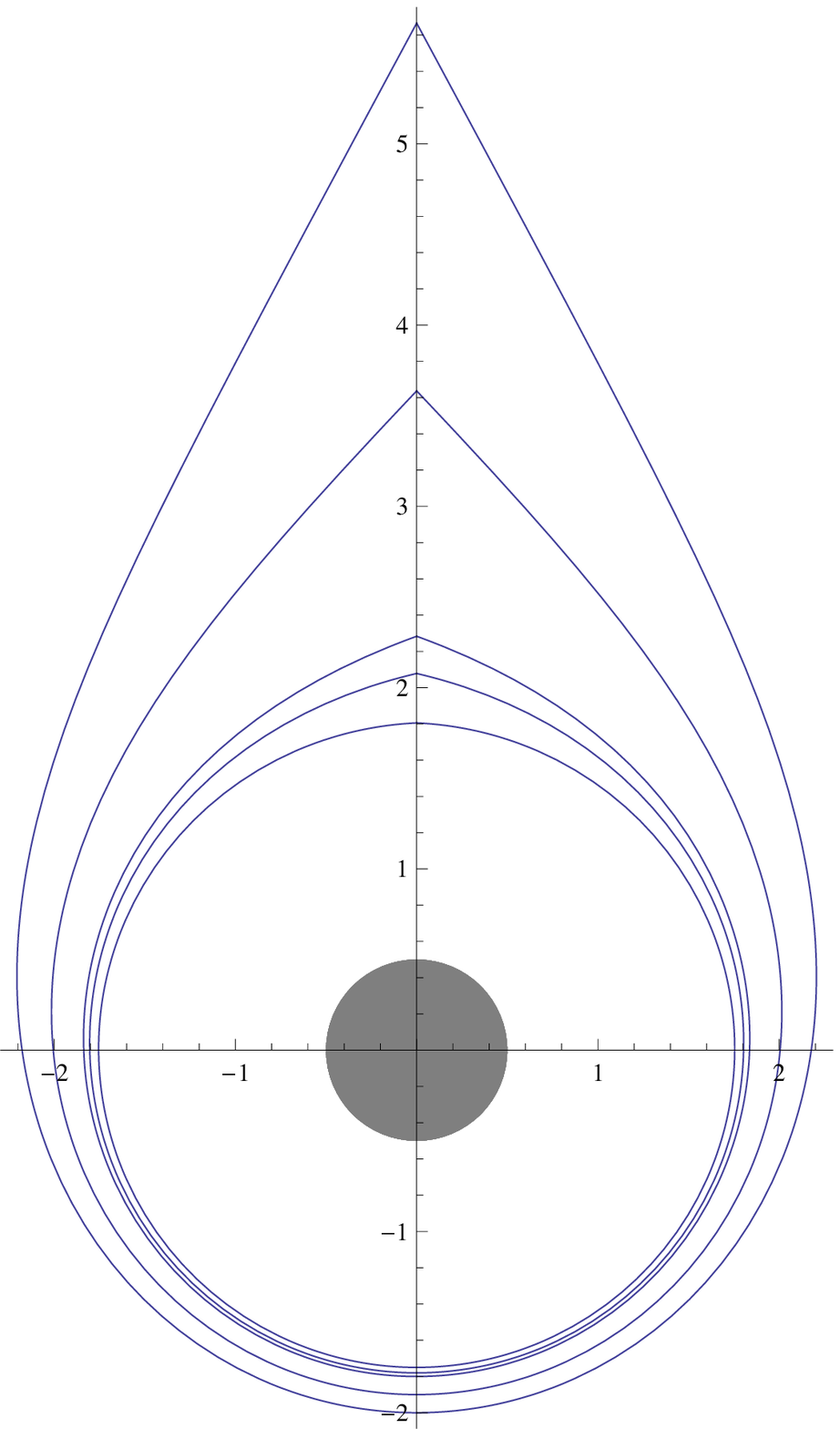}}
\hspace{1cm} 
\subfigure[]{\includegraphics[angle=0,width=0.45 \textwidth]{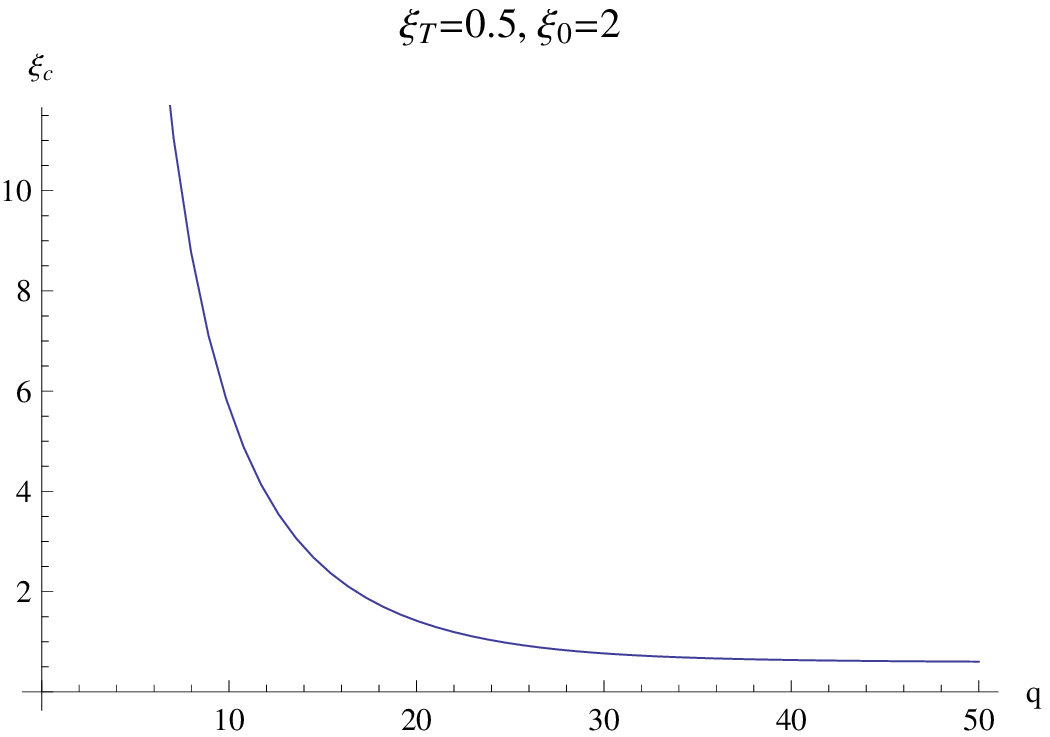}} 
\caption{(a) $\xi_0$ dependence of D5 brane embeddings with $q=10$, $\xi_T =0.5$ . Gray disk denotes to black hole.(b) $q$ dependence of tip of D5 brane($\xi_c$). \label{fig:d5emb}}
\end{center}
\end{figure}

After solving the
equation of motion for gauge field and substituting it to the Lagrangian density, we can get `Hamiltonian' density of D5 brane,
\be\label{Hd5}
{\cal H}_{D5} = \tau_5 
\sqrt{\frac{e^{\Phi}}{2}\frac{\o_-^2}{\o_+}\left(\xi'^2 +\xi^2\right)} \sqrt{\hat{D}(\th)^2+\sin^8
\th}, 
\ee
 where 
\be 
\hat{D}(\th)=-\frac{3}{2}\theta+\frac{3}{2}\sin\theta\cos\theta
+\sin^3\theta\cos\theta.
\ee
Here, we consider all fundamental strings are attached at north
pole. For more detail, see Appendix \ref{BV}.\par

The equation of motion for (\ref{Hd5}) depends on two parameters $q$, $\xi_T$ and  two initial condiation $\xi(\th =0)=\xi_0$, $\xi'(\th=0)=0$. Here we set $\th =0$ is south pole of D5 brane.  For a given value of expectation value of gluon condensation($q$) and temperature($\xi_t$), we can get numerical solution in terms of $\xi_0$. The set of numerical solutions in $\xi$ plane is drawn in Figure
\ref{fig:d5emb}(a). In this figure, a cusp appears at $\th =\pi$ on where $N_c$ fundamental strings are attached.  But these closed D5 brane does not exist in whole value of $q$ and $\xi_t$. When the value of $q$ decrease, the position of tip (we denote it to $\xi_c$) increase and  $q\rightarrow 0$ limit, the tip of D5 brane goes to infinity, \ref{fig:d5emb}(b). It is consistent with the fact that the usual Schwartzschild black hole background does not allow baryon vertex. We also check that the baryon vertex solution does not exist at high temperature.\par
\begin{figure} [ht!]
\begin{center}
\subfigure[]{\includegraphics[angle=0, width=0.45\textwidth]{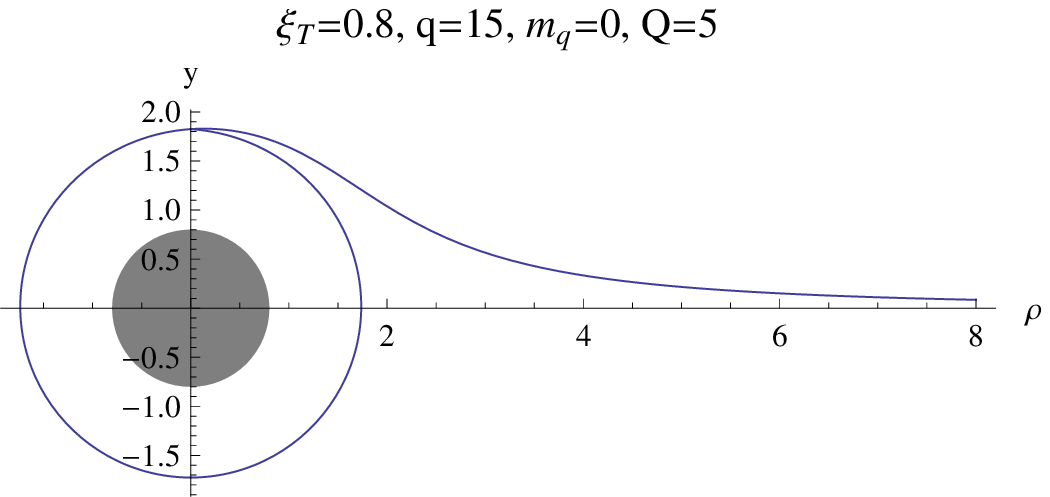}}
\hspace{1cm} 
\subfigure[]{\includegraphics[angle=0,width=0.45\textwidth]{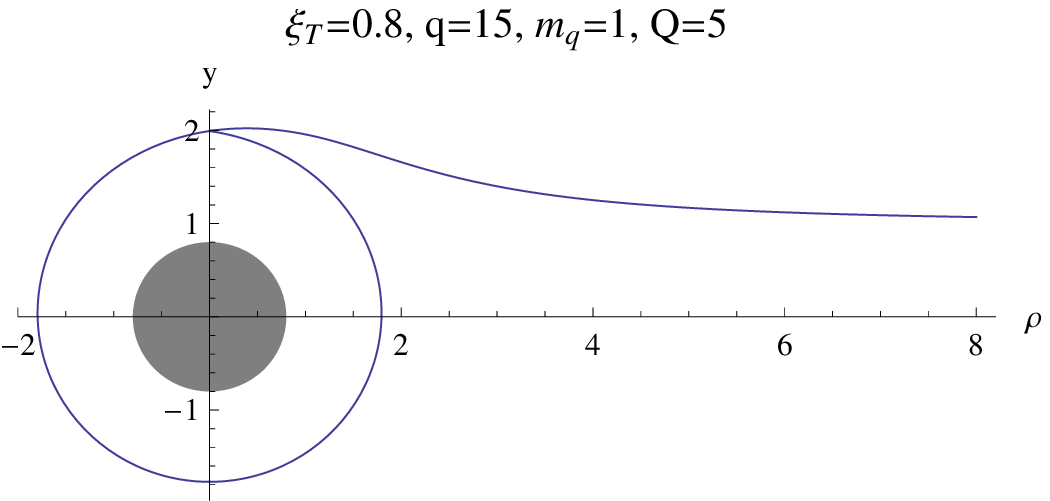}} 
\caption{D7 brane embeddings for $\xi_T =0.8$, $\tilde{Q}=5$ with
(1) $m_q =0$ (b) $m_q=1$  \label{fig:D7Q}}
\end{center}
\end{figure}

As we discussed in \cite{Seo:2008qc}, we can add D7 brane at the tip of D5 brane with force balance condition\cite{Bergman},
\be\label{fbc}
y'(\rho=0) = \frac{\xi_c'}{\xi_c},
\ee
where $\xi_c$ and $\xi_c'$ denote to the position and slop of D5 brane at $\theta=\pi$. For details, see Appendix {\ref{FBC}}.  The full configurations of D7/D5 brane embeddings with certain value of $q$ and density are drawn in Figure \ref{fig:D7Q}. We call this phase as 'baryon phase' because in this phase the physical object is baryon vertex. From this figure, we can see that in the presence of baryon vertex, the slope of probe brane in asymptotic region is always non-zero. 
It means that in baryon phase, chiral symmetry is always broken. 
We found that the embedding corresponding to the baryon phase does not exist at high temperature.

\section{Phase Transition}

As we discussed in previous section, two kinds of  embeddings can exist in  finite temperature.
They correspond to two phases: black hole embedding corresponds to the quark phase and Minkowski embedding 
corresponds to the baryon phase. 
Quark phase can exist whole temperature region while baryon phase  can exist only at low temperature region. 
In our model, the baryon phase exists if temperature is low enough regardless of how high is the density. 
 However, in low temperature, quark phase is also possible. To determine which is the physical phase, we need to compare the free energies of two systems. There are two ensembles we can choose, one is canonical ensemble where density is control parameter. In canonical ensemble we can determine physical phase by comparing free energy. Since we need to determine the configuration 
 in a fixed value of the charge we need  Legendre transformed action which we call `Hamiltonian', although it is not the time translation generator. 
The other one is grand canonical ensemble where chemical potential is control parameter. In this case we have to calculate grand potential which is value of DBI action.  
 
\subsection{Canonical ensemble}
To determine physical solution in canonical ensemble, we have to compare free energies of two phases. Free energy of quark phase is given by integrating Hamiltonian (\ref{HD7}) density for the embeddings,
\be\label{Fbh}
{\cal F}_{\rm quark} (\hat{Q})= \tau_7 \int_{\r_{min}}^{\infty} d\r \hat{{\cal H}}_{D7}(\hat{Q})\Big|_{\rm quark~phase},
\ee
where $\hat{{\cal H}}_{D7}={\cal H}_{D7}/\tau_7$ (we introduce it for convenience).
Here, we regularized free energy by subtracting energy of flat D7 brane.\par

On the other hand, baryon phase needs compact D5 brane, called baryon vertex operator. 
Since each D5 should have $N_c$ quarks, their number is related to the quark number by $ N_B =Q/N_c$.  
The total free energy in baryon phase  can be obtained by  adding energy of D5 brane to that of D7 brane:
\bea\label{Fbary}
{\cal F}_{\rm baryon} (\hat{Q}) &=& \tau_7 \int d\r \hat{{\cal H}}_{D7}(Q)\Big|_{\rm baryon~phase} + \frac{Q}{N_C} \tau_5 \int d\th \hat{{\cal H}}_{D5} \cr\cr
&=& \tau_7 \left[ \int d\r \hat{{\cal H}}_{D7}(\hat{Q})\Big|_{\rm baryon~phase} + \frac{2}{3 \pi} \hat{Q} \int d \th\hat{{\cal H}}_{D5}\right].
\eea

\begin{figure} [ht!]
\begin{center}
\subfigure[]{\includegraphics[angle=0, width=0.3\textwidth]{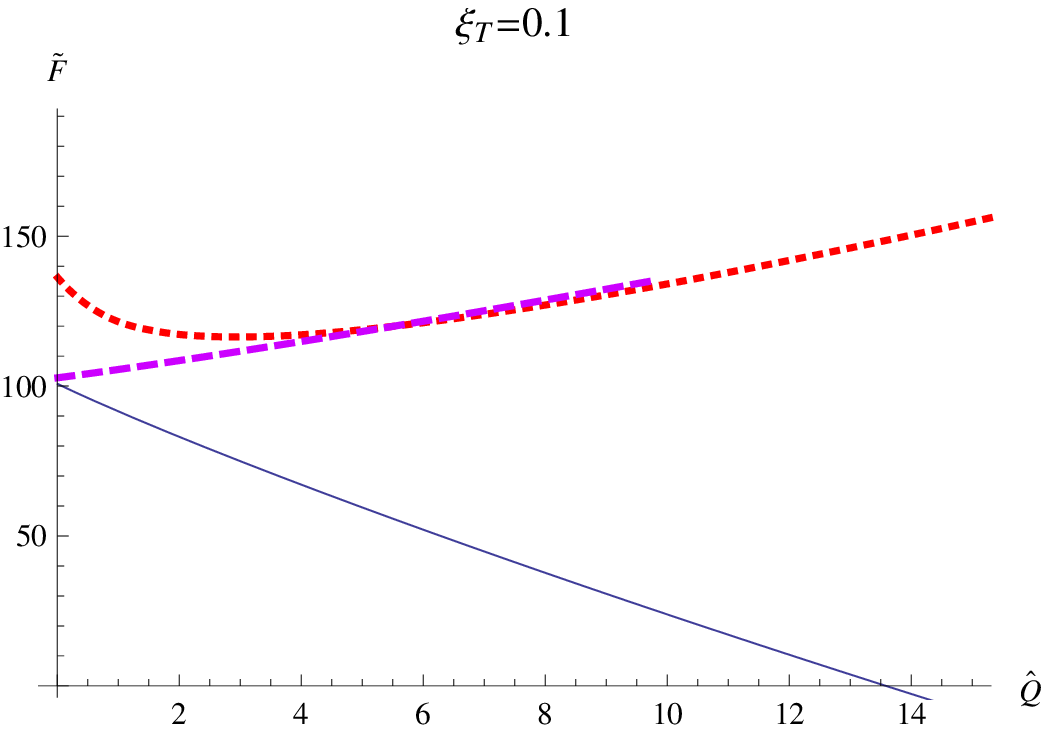}}
\hspace{1cm}
\subfigure[]{\includegraphics[angle=0,width=0.3\textwidth]{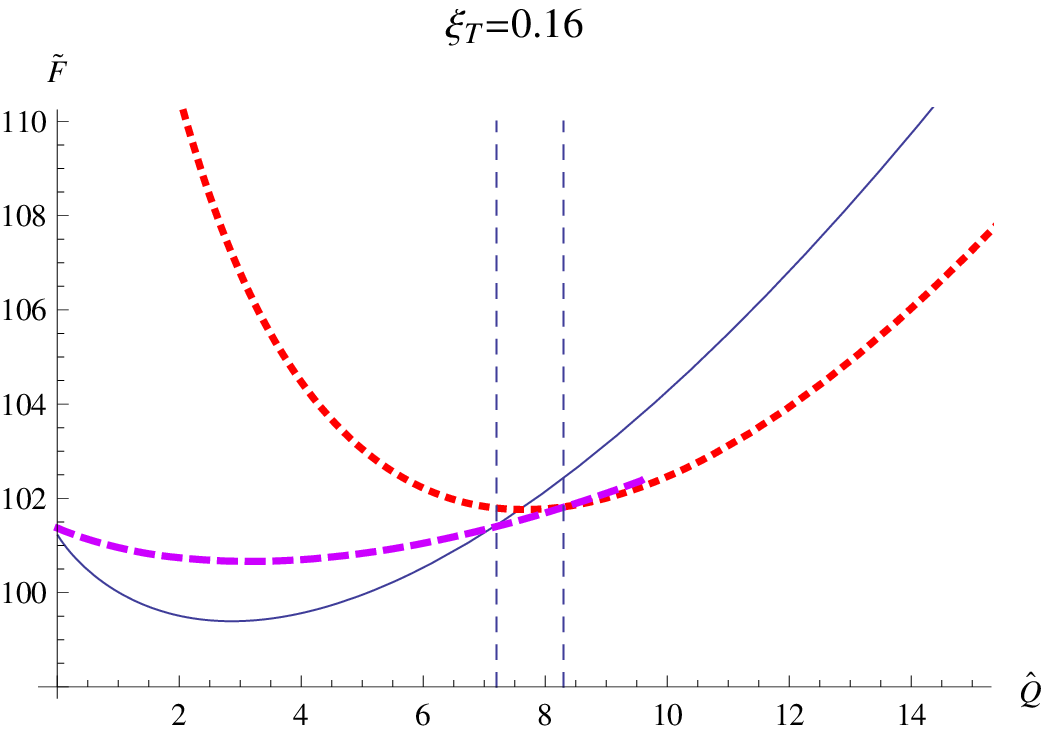}}
\hspace{1cm}
\subfigure[]{\includegraphics[angle=0,width=0.3\textwidth]{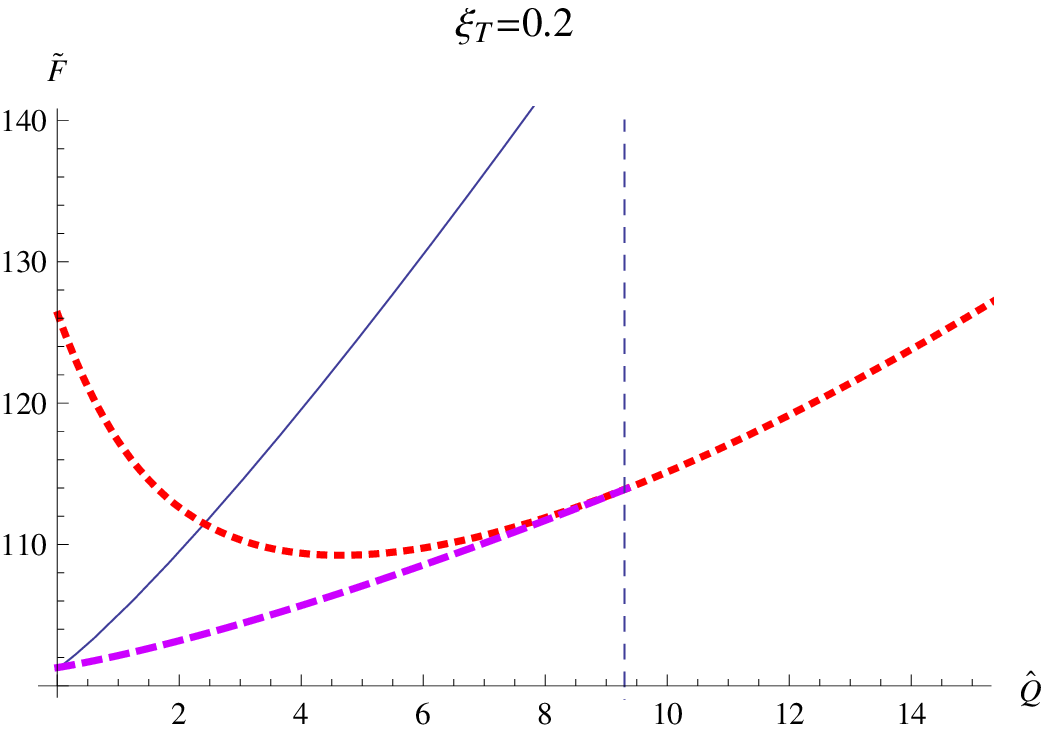}}
\hspace{1cm}
\subfigure[]{\includegraphics[angle=0,width=0.3\textwidth]{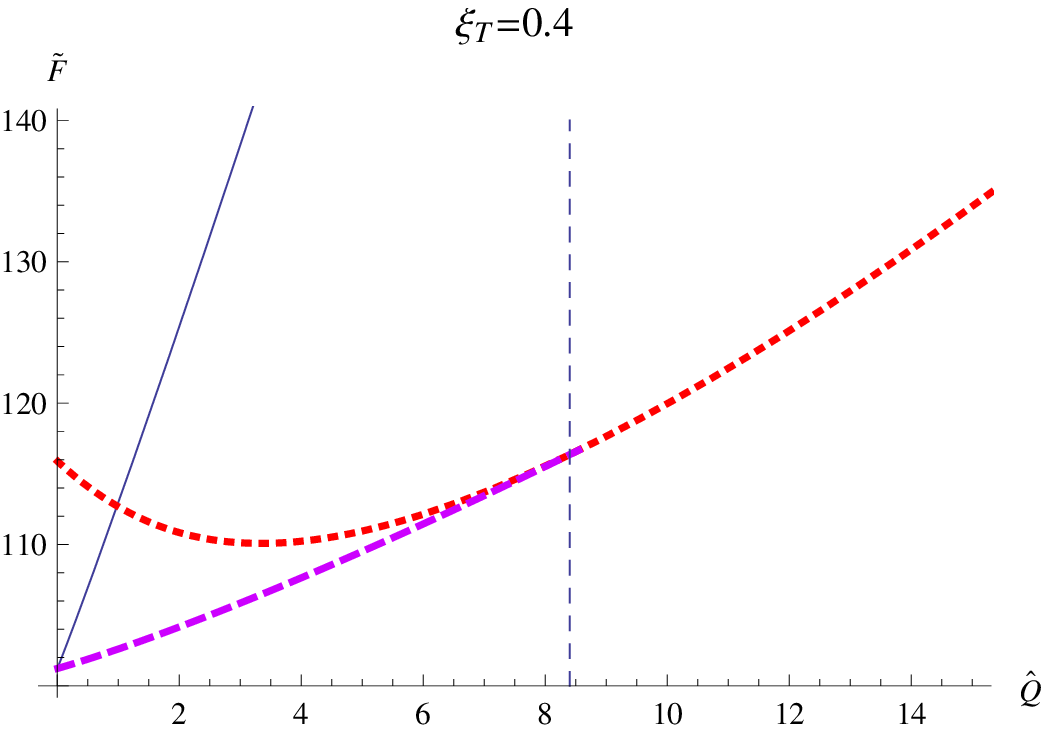}}
\caption{Density dependence of free energy for massless quark case with $q=15$. To visualize the difference of free energy, we plotted $\tilde F$ which is defined as $\tilde F=F-\alpha(T) Q$ for some $\alpha(T)$.  
Solid line is for baryon phase, 
 red dotted line is for  chiral symmetry restored quark phase and dashed purple line is for quark phase with chiral symmetry broken. Vertical dotted line denotes  phase transition point.   
\label{fig:FQ}}					
\end{center}
\end{figure}

By comparing the value of (\ref{Fbh}) and (\ref{Fbary}), we can determine which phase is physically favored.
As we discussed in previous section, there are two phases in quark phase, which is true also  in massless quark case. 
One is quark phase with broken chiral symmetry and the other one is phase with chiral symmetry restored. 
The density dependences of free energy in massless quark case are drawn in Figure \ref{fig:FQ}.
In the figure,  we plotted $\tilde F$ which is defined as $\tilde F=F-\alpha(T) Q$ to visualize the difference of the 
free energies of different phases. 
One should notice that  all the free energies   monotonically increase as function of density, 
although the figure does not show that due to the subtraction of $\alpha(T)Q$. 
 
\begin{figure} [ht!]
\begin{center}
\subfigure[]{\includegraphics[angle=0,width=0.4\textwidth]{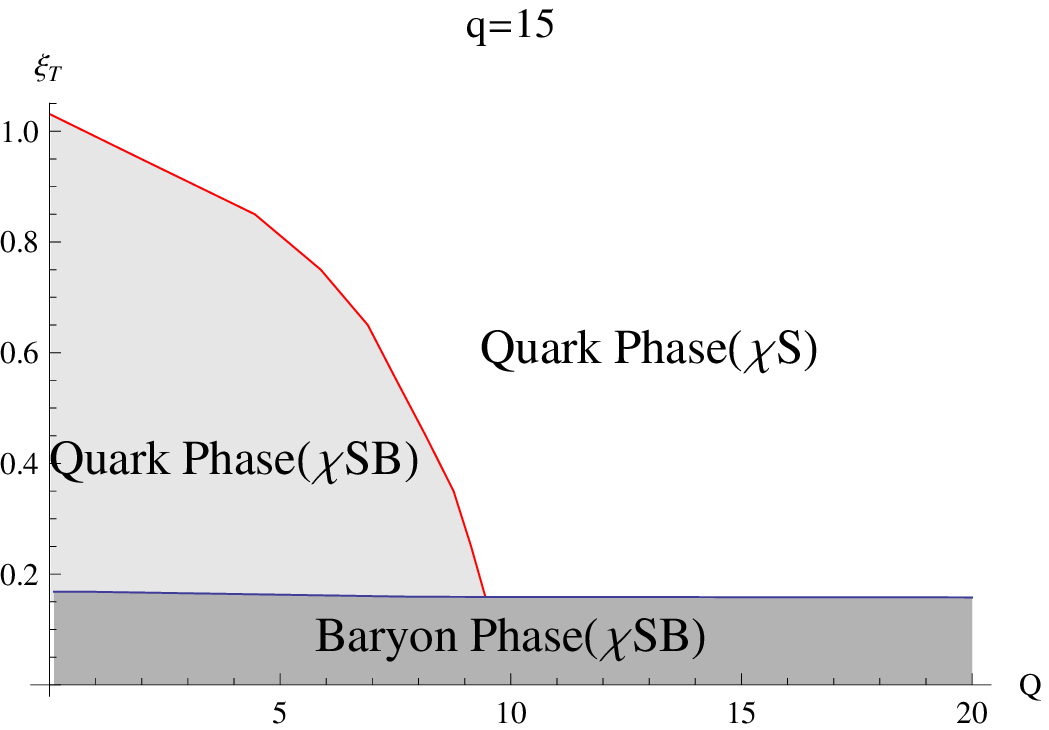}}
\hspace{1cm}
\subfigure[]{\includegraphics[angle=0,width=0.4\textwidth]{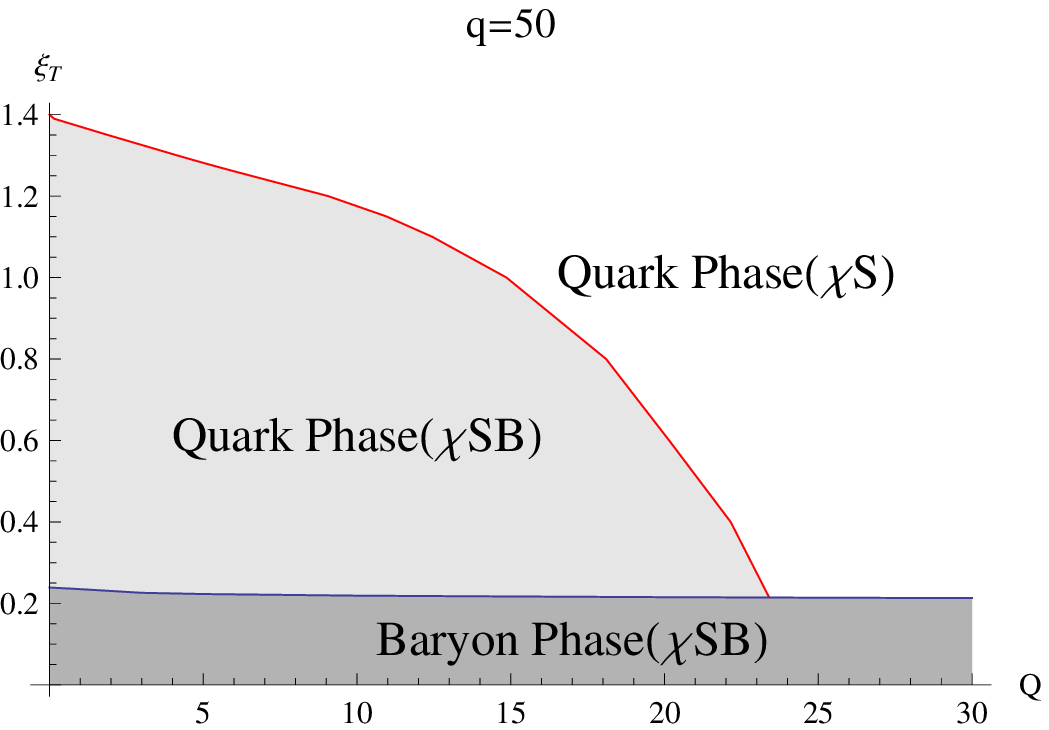}}
\caption{Density dependence of phase transition temperature between quark phase and baryon phase (a) with $q=15$, (b)with $q=50$.
\label{fig:PT_mq0}}
\end{center}
\end{figure}

At low temperature, the free energy of baryon phase is always lower than that of quark phase for all density region, see Figure \ref{fig:FQ}(a). It means that the baryon phase is physical at low temperature. As temperature increase, the free energy lines  change drastically, see Figure \ref{fig:FQ}(b). At  temperature $\xi_t =0.16$, the free energy of baryon phase is lowest in  low density region. As density increase, there is phase transition between baryon phase and quark phase with broken chiral symmetry(first vertical line in \ref{fig:FQ}(b)). After the baryon to quark phase transition, there is another phase transition between quark phases: 
from  broken   to restored chiral symmetry quark phase. At higher temperature, the free energy of quark phase is always smaller than that of baryon phase.  But there is a phase transition from chiral symmetry broken phase to chiral symmetry restored phase at certain density. See  Figure \ref{fig:FQ}(c), (d). The chiral phase transition between two quark phases coincide with the result of previous section, Figure \ref{fig:chiralbds}. Of course, at high enough temperature, the chiral symmetry restored quark phase is the only  physical phase.  
These are summarized in the phase diagram drawn in Figure \ref{fig:PT_mq0}.

This phase diagram has rich structure. At low temperature, baryon phase is always physical one in all density region. And chiral symmetry is always broken in this phase. However, as we increase temperature, phase transition appears differently depends on density. At low density, as temperature increase, baryon phase changes into quark phase but chiral symmetry is still broken. If we increase temperature more, then there is chiral symmetry restoration transition in quark phase. On the other hand, at high density, as we increase temperature, baryon/quark phase transition and chiral phase transition appears at the same time. \par
We   find that the baryon/quark phase transition temperature decreases as density increase. However, the decreasing rate is too slow  so that phase transition temperature looks like constant in the figure. The value of $q$ also affect phase transition. As $q$ increase,  both phase boundaries move up to larger temperature and larger density region maintaining the overall shape. See Figure \ref{fig:PT_mq0}(b).  
Notice that our phase diagram is very similar to the ones for Sakai-Sugimoto model studied in  \cite{Aharony:2006da}, where 
the baryon phase should be replaced by the soliton geometry.  
 Notice that in our model, there is no Hawking Page transition since there is no scale other than the temperature
and the dilaton contribution is precisely canceled by that of the axion. 
 Our baryon phase is still in the black hole geometry therefore the dynamic origin is very different.
 
\vspace{0.2cm}

\begin{figure} [ht!]
\begin{center}
\includegraphics[angle=0,width=0.4\textwidth]{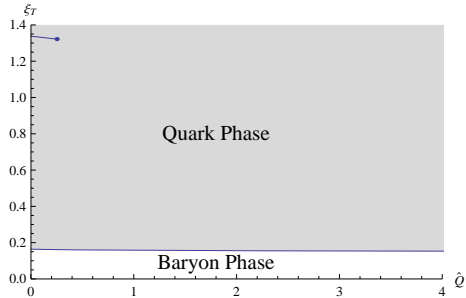}
\caption{Density dependence of phase transition temperature between quark phase and baryon phase for $m_q =1$ and $q=20$.
\label{fig:PT01}}
\end{center}
\end{figure}

In the case of $m_q \ne 0$, chiral symmetry is explicitly  broken so that  we might  expect that there is only baryon/quark phase transition.  However, as we discussed in previous section, there is another phase transition in quark phase.  
Look at the short line at the up-left region of  the phase diagram in Figure \ref{fig:PT01}. It is for $m_q =1$.

As we decreases the quark mass, the end point of the phase boundary  
extends to higher density and lower temperature region so that when $m_q\to 0$, we recover the phase diagram drawn in Figure \ref{fig:PT_mq0}. Actually the line becomes long very fast  when $m_q$ gets to the near zero value.  
Contrarily for $q=0$,  the line for the  transition from black hole embedding (BHE) to  another BHE  exist only in very small density region and it disappear when $m_q$ goes to zero. 
Therefore  this is very characteristic feature caused by the gluon condensation $q$.

\subsection{Relation between chemical potential  and density}
Equations of states which are the relations between thermodynamically conjugate variables like  chemical potential/density, quark mass/chiral condensation e.t.c. play important role to understand phase structure. In this section, we will discuss the equation of state describing the relation between chemical potential and density. Thermodynamically chemical potential is defined as derivative of free energy with respect to density,
\be
\mu =\frac{\partial{\cal F}}{\partial Q}.
\ee
In quark phase, we can get
\bea\label{muquark}
\mu_{quark}&=& \frac{\partial{\cal F}_{quark}}{\partial Q}\cr
&=& \frac{1}{2 \pi \a'} \int_{\r_{min}}^{\infty} \frac{\partial \hat{{\cal H}}_{D7}}{\partial \hat{Q}} \cr
&=& \int_{\r_{min}}^{\infty} d\r\, \partial_{\r} A_t = A_t (\infty)-A_t(\r_{min}).
\eea
In this case, $A_t(\r_{min}) =0$ because $\r_{min}$ is black hole horizon. It is consistant with usual definition of chemical potential(tale of $A_t$ field) from AdS/CFT correspondence. On the other hand, the free energy of baryon phase contains mass of baryon vertex, chemical potential should have mass of source.
\bea
\label{mubaryon}
\mu_{baryon} &=& \frac{\partial{\cal F}_{baryon}}{\partial Q}\cr
&=&\frac{1}{2 \pi \a'} \int_{\r_{min}}^{\infty} \frac{\partial \hat{{\cal H}}_{D7}}{\partial \hat{Q}} + \frac{1}{2\pi\a'}\frac{2}{3 \pi}  \int d \th\hat{{\cal H}}_{D5} \cr
&=& A_t(\infty)-A_t(\r_{min}) +\frac{1}{N_c}\int d\th {\cal H}_{D5}.
\eea
The last term in (\ref{mubaryon}) the  baryon mass divided by $N_c$,  which is the constituent quark mass. 
 { Therefore, the chemical potential contains mass of the source. In quark phase, the source is the fundamental strings. 
 For the blackhole embedding,  D7 brane touches  the black hole (BH) horizon and  the fundamental strings are replaced by deformation of D7 brane. Therefore  the energy of fundamental strings is contained 
 in  that of D7 brane. That's why there is no source term, the analogue of the last term in the eq.(\ref{mubaryon})),  in the chemical potential of the quark phase. 
The density dependences of chemical potential for $m_q=0$ embedding are drawn in Figure \ref{fig:muq00}.  We can see that chemical potentials    monotonically increase as density increases for any phases. Therefore,  there is no thermodynamical instability in $m_q =0$ case. 
 
\begin{figure} [ht!]
\begin{center}
\subfigure[]{\includegraphics[angle=0, width=0.35\textwidth]{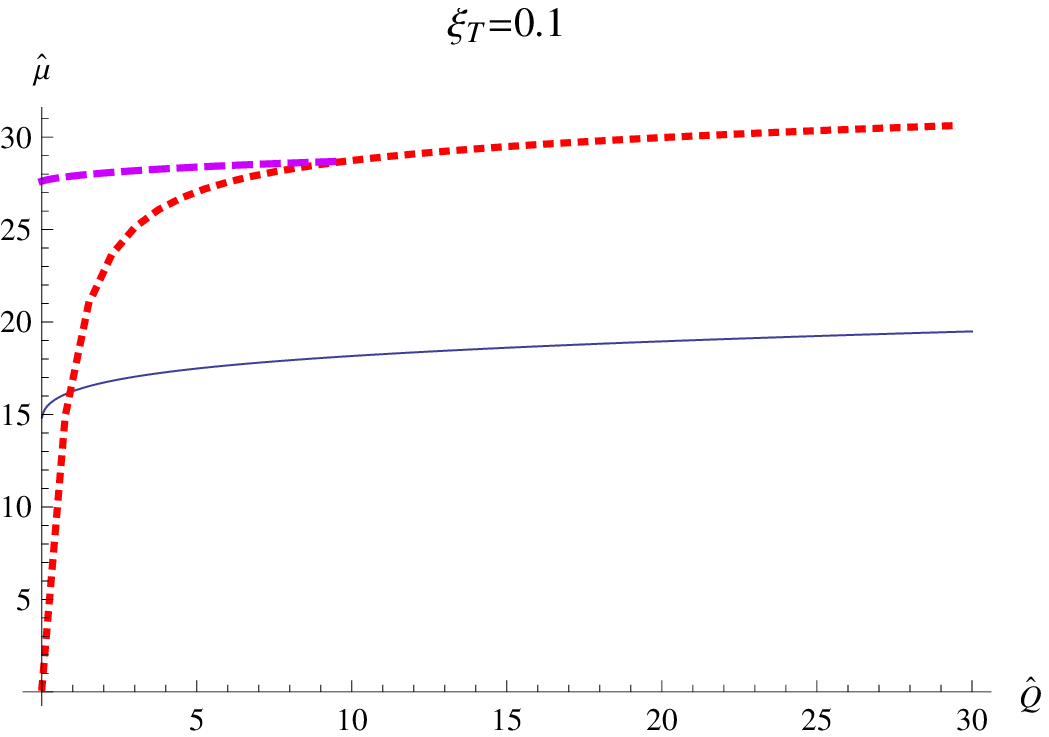}}
\hspace{1cm}
\subfigure[]{\includegraphics[angle=0,width=0.35\textwidth]{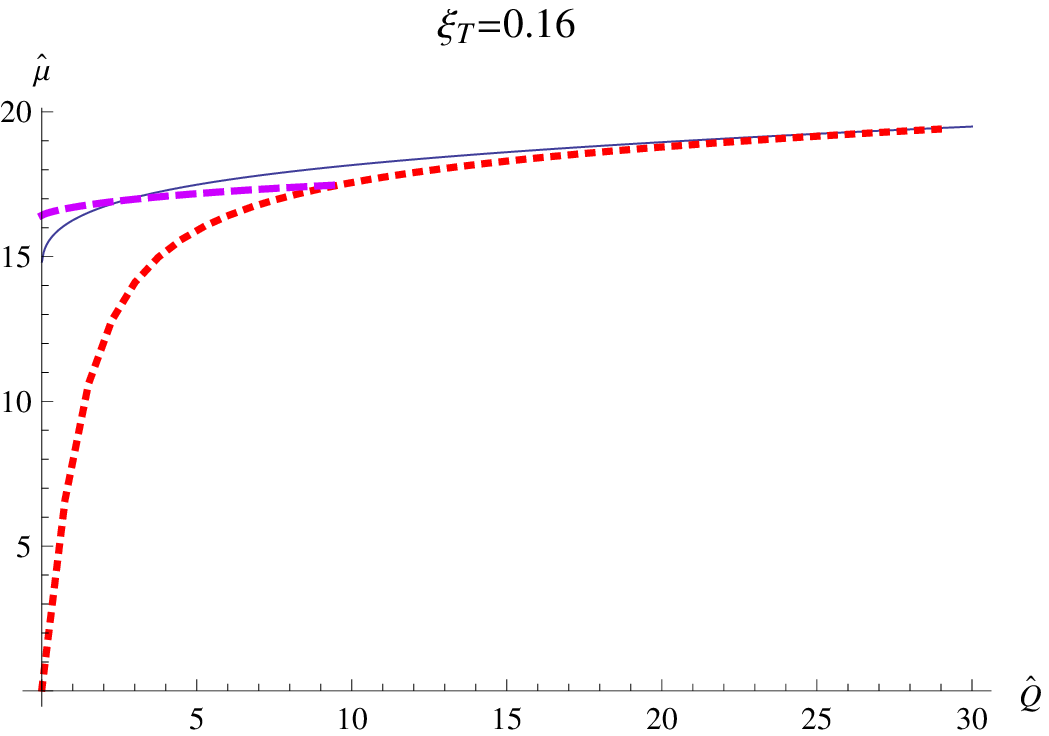}}
\subfigure[]{\includegraphics[angle=0,width=0.35\textwidth]{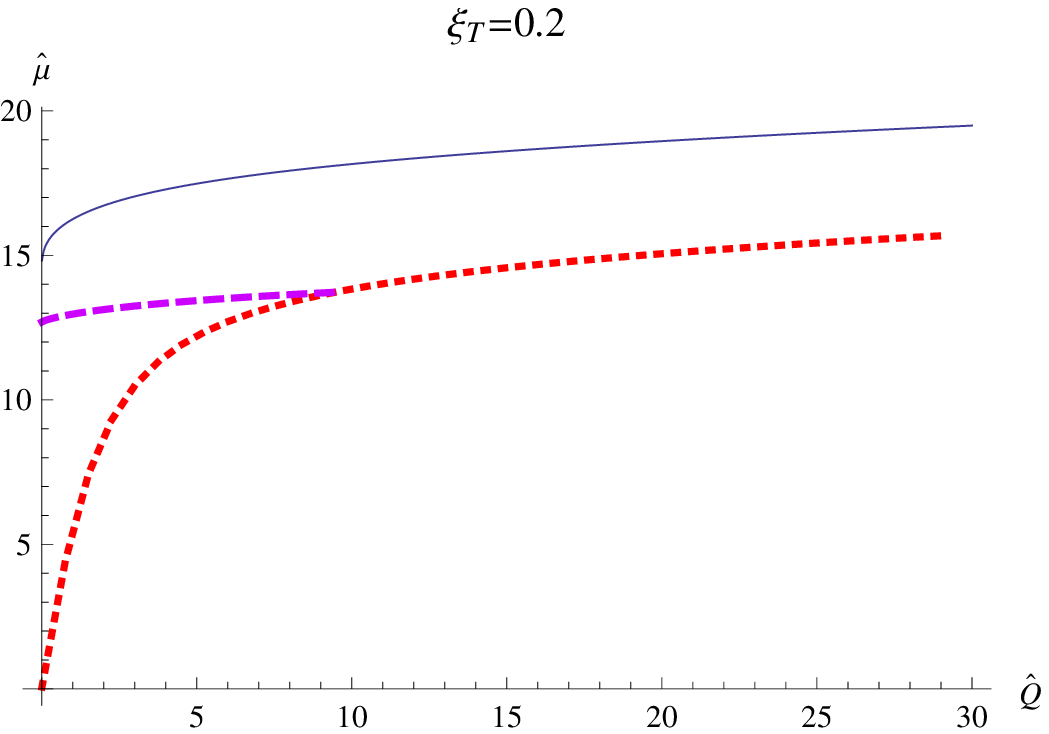}}
\hspace{1cm}
\subfigure[]{\includegraphics[angle=0,width=0.35\textwidth]{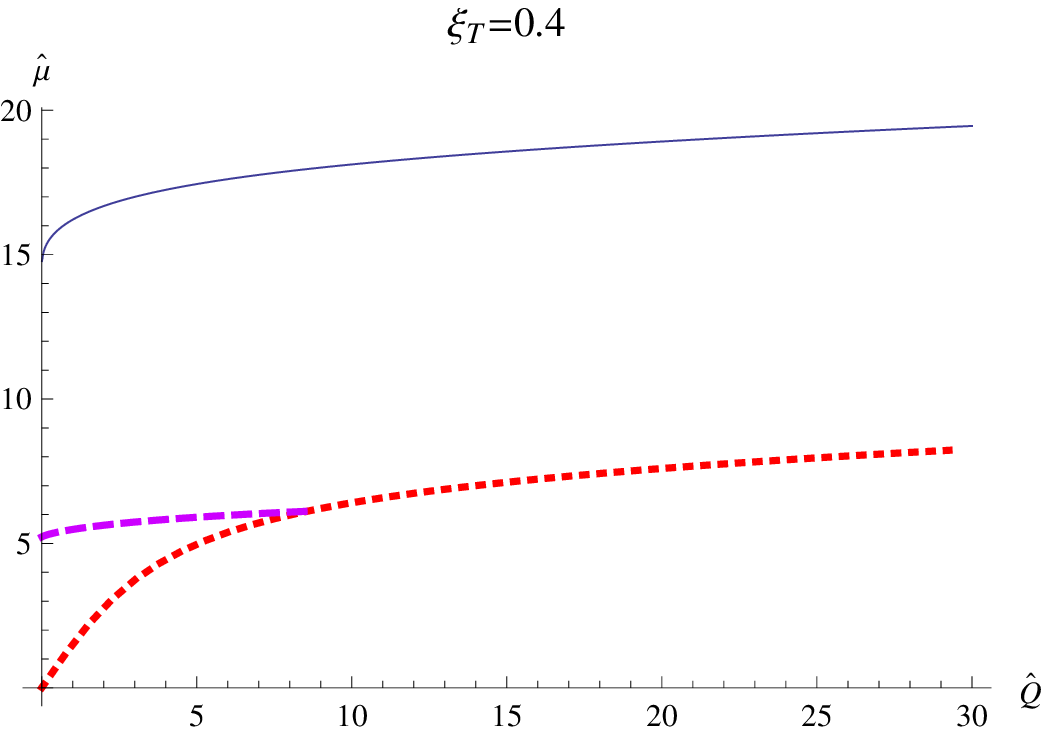}}
\caption{Density dependence of chemical potential for $m_q=0$. Red dotted line denotes chemical potential for quark phase with chiral symmetry and purple dashed line is for quark phase with broken chiral symmethry.  The solid line denotes  baryon phase.\label{fig:muq00}}
\end{center}
\end{figure}


\begin{figure} [ht!]
\begin{center}
\subfigure[]{\includegraphics[angle=0,width=0.4\textwidth]{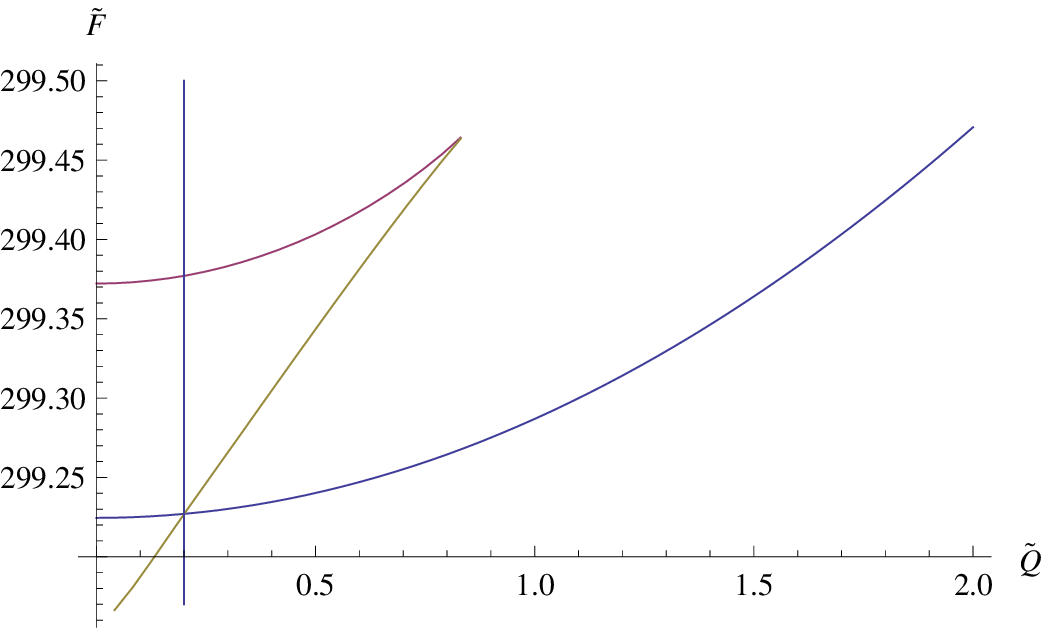}}
\hspace{1cm}
\subfigure[]{\includegraphics[angle=0,width=0.4\textwidth]{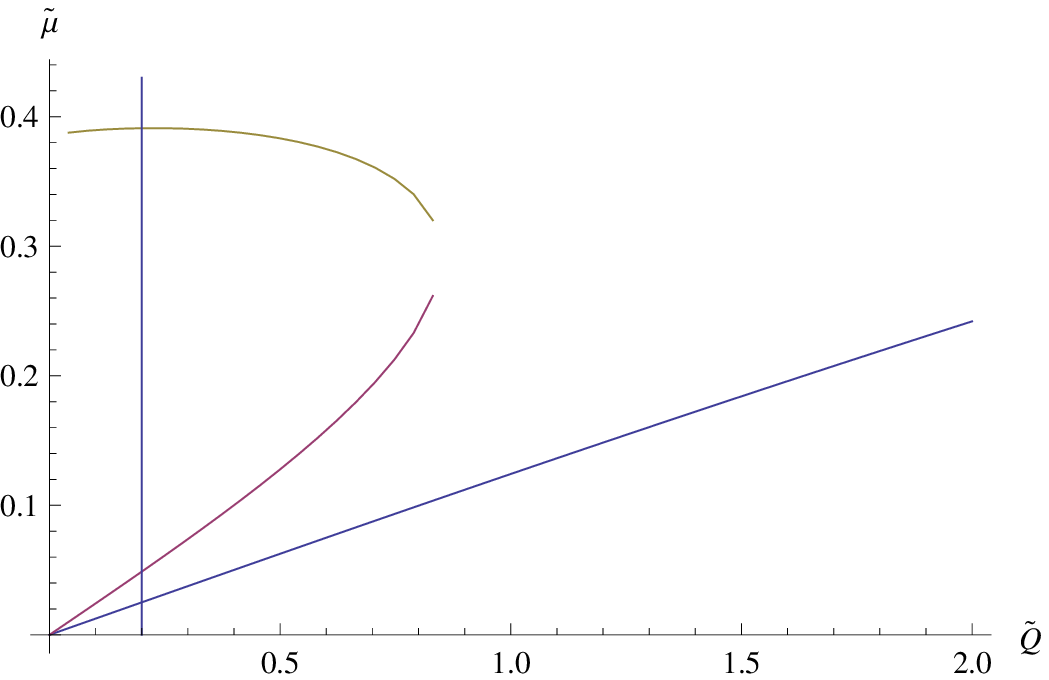}}
\caption{(a) Free energy as a function of density in certain temperature for   $m_q=1$. 
(b) Density dependence of chemical potential, $\tilde{\mu}=2\pi\a' \mu$.  In both figure $\xi_T=1.3$,  $q=50$.   \label{fig:muq01}}
\end{center}
\end{figure}

The free energy and chemical potential with finite quark mass as  functions of density  are drawn in Figure \ref{fig:muq01}. 
Figure \ref{fig:muq01}(a) shows density dependence of free energy at finite temperature with $q=50$ and $m_q=1$. At certain density there is phase transition in quark phase. In usual  AdS Schwartzschild background without $q$, there is a thermodynamical instability associated with the negative slope branch of the  $\mu$-$Q$ diagram. But in Figure \ref{fig:muq01}(b), we can see that chemical potential increases at low density and phase transition happens before chemical potential
begin to  decrease. Therefore, chemical potential for physical state  always  increase monotonically as density increase($\partial \mu/\partial Q >0$),  hence there is no thermodynamical instability.\par 
As temperature decreases, both quark phase and baryon phase exist and the shape of chemical potential is monotonic as function of the density in both phases which is similar to Figure \ref{fig:muq00}.

\subsection{Grand canonical ensemble}
In this section, we will discuss the system with grand canonical ensemble. Grand potential is thermodynamically defined as
\be\label{grand}
\Omega = {\cal F} -\mu Q.
\ee
From definition of `Hamiltonian density and number density, grand potential for quark phase can be written as
\bea\label{gquark}
\Omega_{quark}&=& {\cal F}_{quark}-\mu_{quark} Q \cr\cr
&=& \int \tilde{F}\frac{\partial {\cal L}_{D7}}{\partial \tilde{F}} \Big|_{quark~phase}-\int {\cal L}_{D7}|_{quark~phase} -\int {\tilde F}{\tilde Q}\cr 
&=& -\int {\cal L}_{D7}|_{quark~phase}.
\eea
The last term in (\ref{gquark}) is nothing but value of on-shell DBI action for black hole phase. It it obvious because `Hamiltonian density' is defined by Legendre transformation of DBI action. 
Grand potential for baryon phase becomes
\bea\label{gbaryon}
\Omega_{baryon} &=& {\cal F}_{baryon} -\mu_{baryon} Q\cr\cr
&=& \int {\cal H}_{D7}|_{baryon~phase} +\frac{Q}{N_c}\int {\cal H}_{D5} -\left(\int \tilde{F}\tilde{Q}+\frac{Q}{N_c}\int {\cal H}_{D5}\right)\cr
&=& -\int {\cal L}_{D7}|_{baryon~phase}.
\eea
We can see that energy of source in free energy canceled by one in chemical potential, finial form of grand potential is value of DBI action of D7 brane for baryon phase. D5 part does not contribute to grand potential. 

\begin{figure} [ht!]
\begin{center}
\subfigure[]{\includegraphics[angle=0, width=0.35\textwidth]{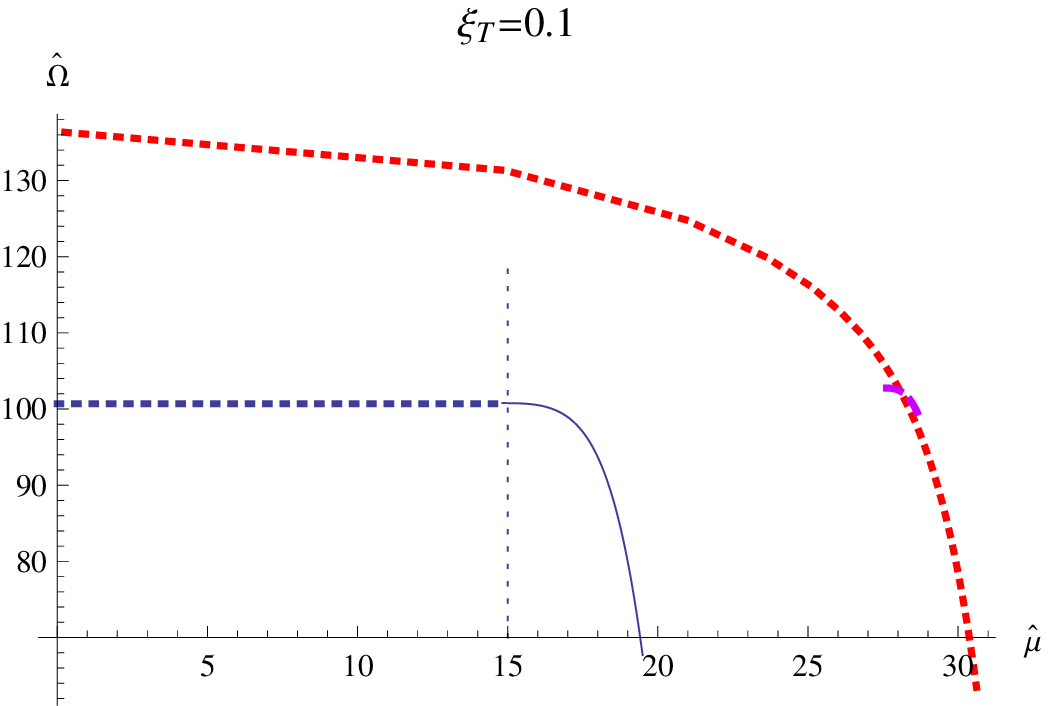}}
\hspace{1cm}
\subfigure[]{\includegraphics[angle=0,width=0.35\textwidth]{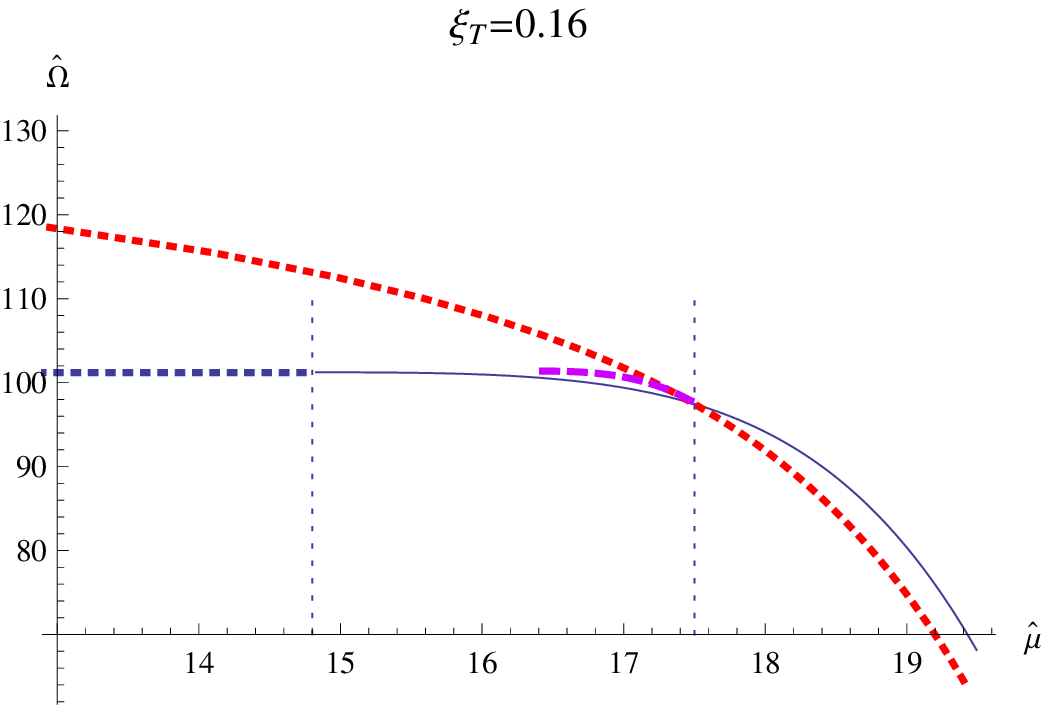}}
\subfigure[]{\includegraphics[angle=0,width=0.35\textwidth]{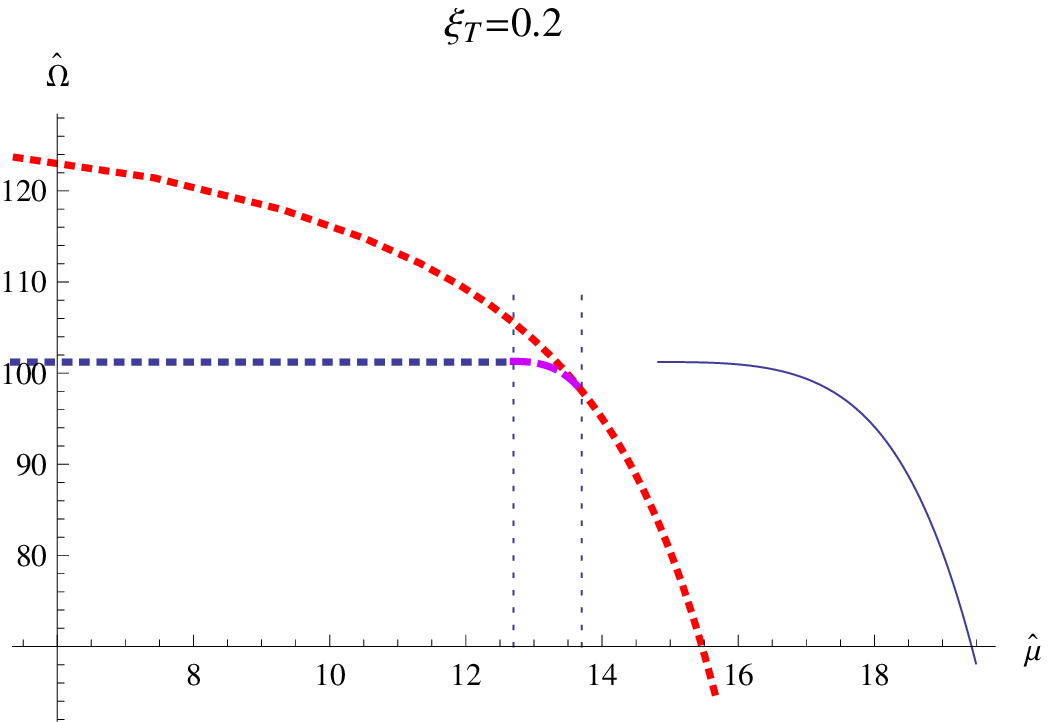}}
\hspace{1cm}
\subfigure[]{\includegraphics[angle=0,width=0.35\textwidth]{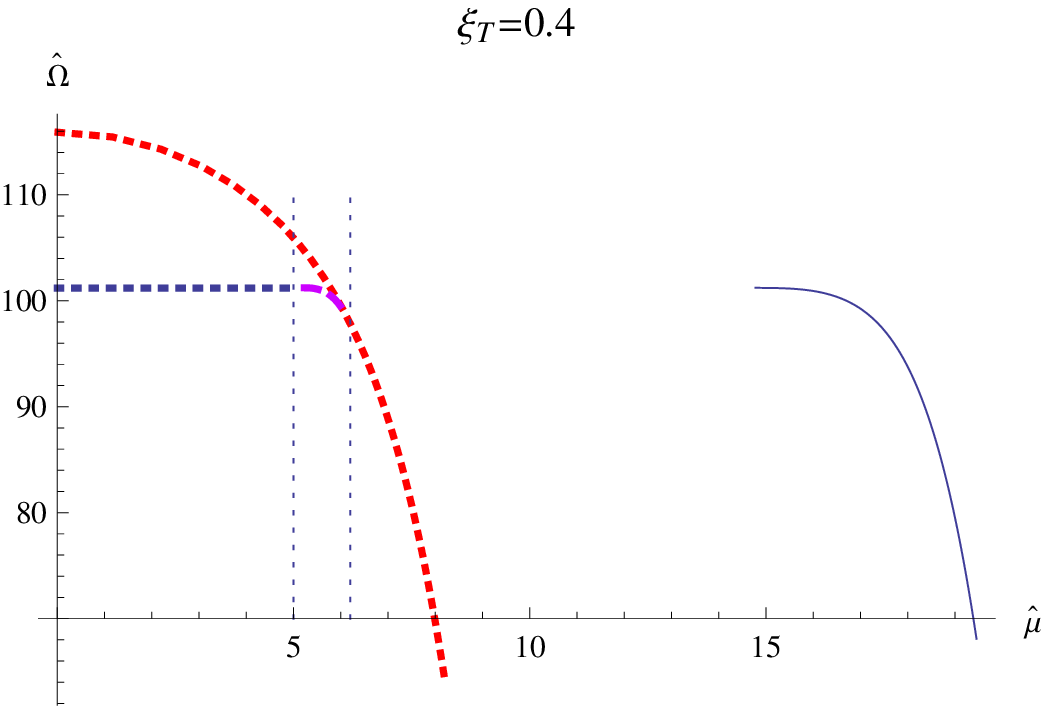}}
\caption{Chemical potential dependence of grand potential($m_q =0$ and $q=15$). Red dotted line denotes to grand potential for chiral symmetry restored quark phase purple dashed line to grand potential for quark phase without chiral symmetry and solid line to baryon phase \label{fig:gmu00}}
\end{center}
\end{figure}

In the massless quark case($m_q =0$), the grand potentials of quark and baryon phase are drawn in Figure \ref{fig:gmu00}.
In each case of the that figure, chemical potential start from non-zero value as one can see in Figure \ref{fig:muq00}.
 It implies that if the chemical potential is not large enough to the energy of particle, no particle can be created. Therefore, the system remains as vacuum until chemical potential reach to the energy of particle in the system. 
 Therefore the value of the chemical potential  should  be identified as the constituent quark mass. 
  Naturally the difference in the chemical potential in baryon phase and  that in the realized quark phase 
(between the two quark phases) can be identified as the binding energy. 
The   density dependence of the   quark mass is plot in figure 19. 
The phase transition point can be identified precisely as the point the binding energy is 0. 
This point is of course the point where baryons melt. The melting point in temperature-density plane is nothing but our phase diagram.

 From the figure  Figure \ref{fig:gmu00},  one can easily read off the  critical  chemical potential value 
 points as one increases the temperature. 

\begin{figure} [ht!]
\begin{center}
\subfigure[]{\includegraphics[angle=0,width=0.45\textwidth]{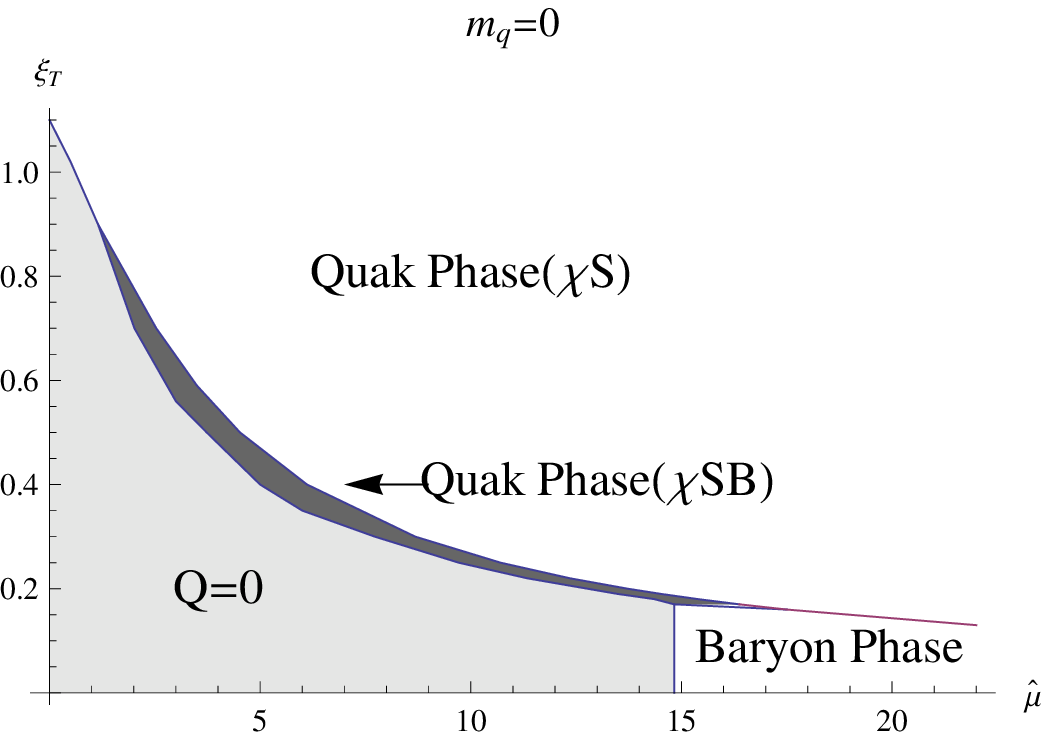}}
\subfigure[]{\includegraphics[angle=0,width=0.45\textwidth]{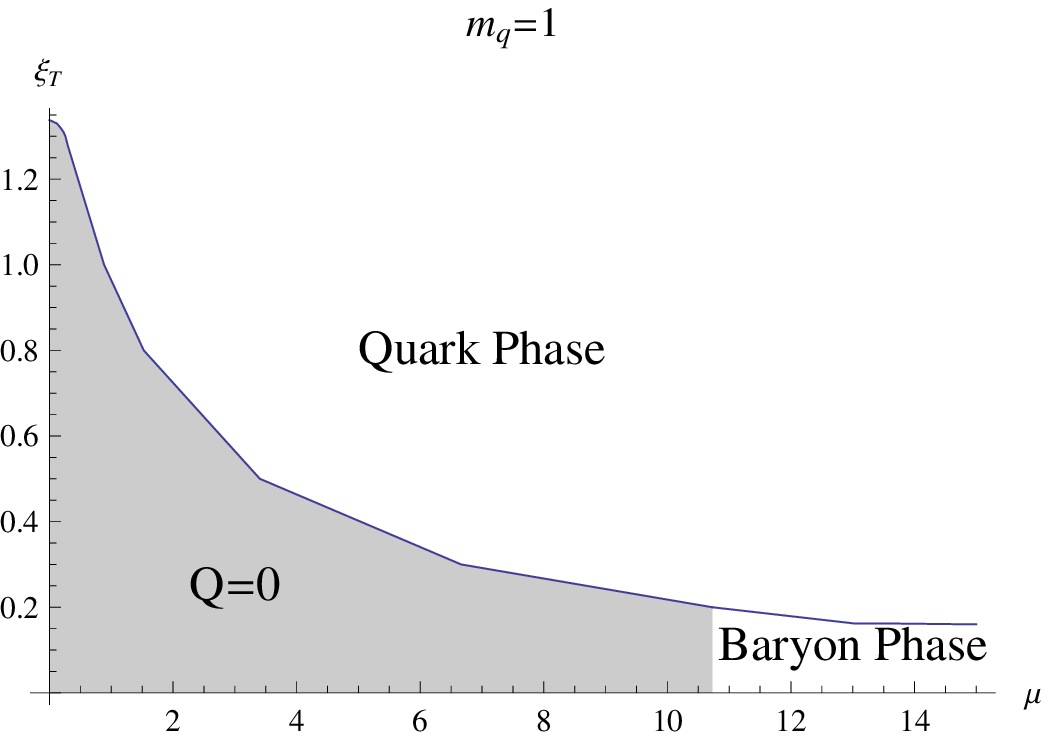}}
\caption{ Phase boundary in grand canonical ensemble in the case of (a) $m_q =0$, (b) $m_q =1$.
\label{fig:PTgrand00}}
\end{center}
\end{figure}

In the case of finite quark mass, chiral symmetry is always broken similar to the canonical ensemble. We have only two phases, the baryon phase and quark phase. However, we find that the density dependence of chemical potential and the chemical potential dependence of grand potential behave similarly to the massless case except the absence of chiral phase transition.
The phase structure based on above discussion is drawn in Figure \ref{fig:PTgrand00}. Similar to phase diagram in canonical ensemble, the temperature of phase transition between quark and baryon phase is almost constant. We also find the transition in quark phase in small density region, but this line is not shown in this diagram.
 
\subsection{Chiral condensation}
One of the most important observable in QCD is the chiral condensation (CC) and its density dependence. 
It measures the dynamical mechanism for creating the mass of the hadrons. 
Figure  \ref{fig:gmu01} shows the density dependence of the chiral condensation. 
In the baryon phase, chiral symmetry is always broken and there is a quark phase where chiral symmetry is broken. 
Our result shows that CC is increasing in baryon phase and decreasing in the relevant quark phase. 

\begin{figure} [ht!]
\begin{center}
\subfigure[]{\includegraphics[angle=0, width=0.35\textwidth]{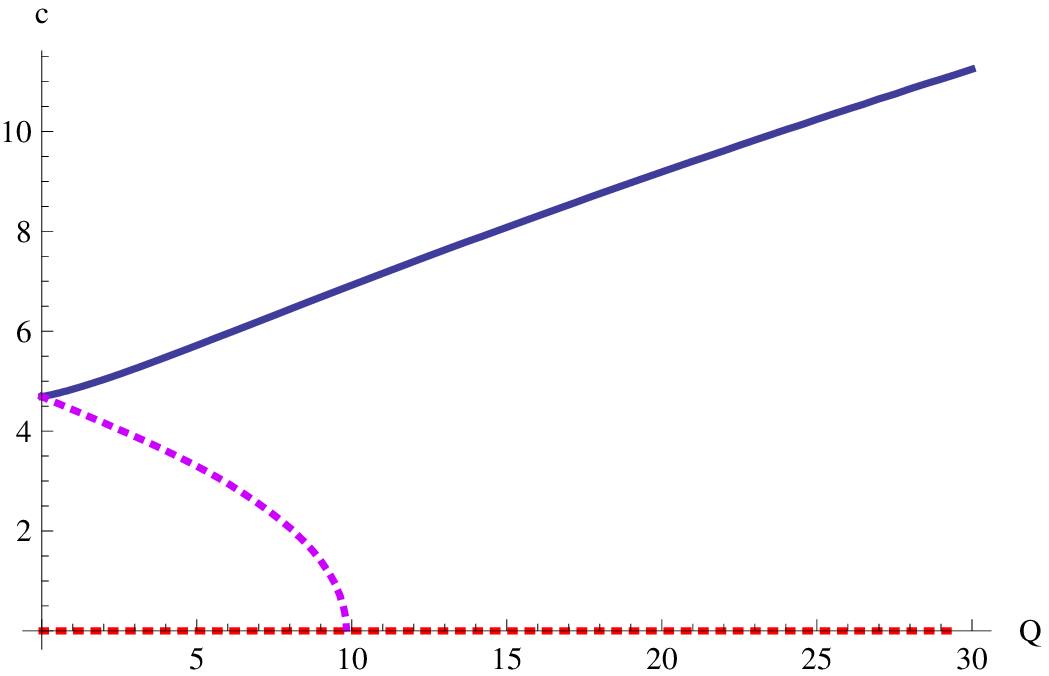}}
\hspace{1cm}
\subfigure[]{\includegraphics[angle=0,width=0.35\textwidth]{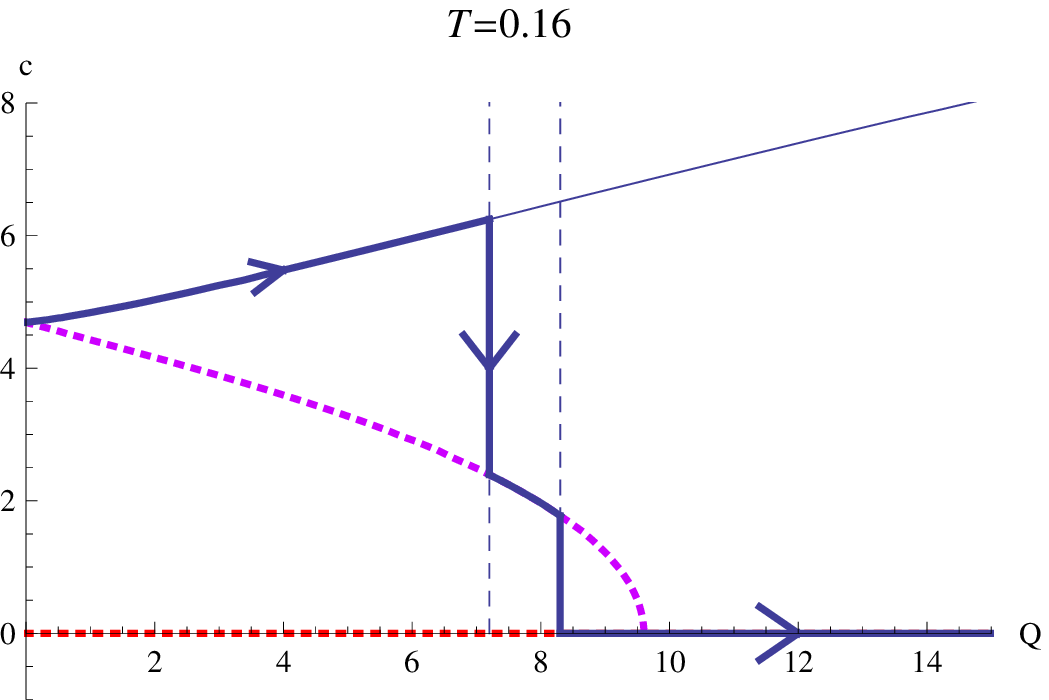}}
\subfigure[]{\includegraphics[angle=0,width=0.35\textwidth]{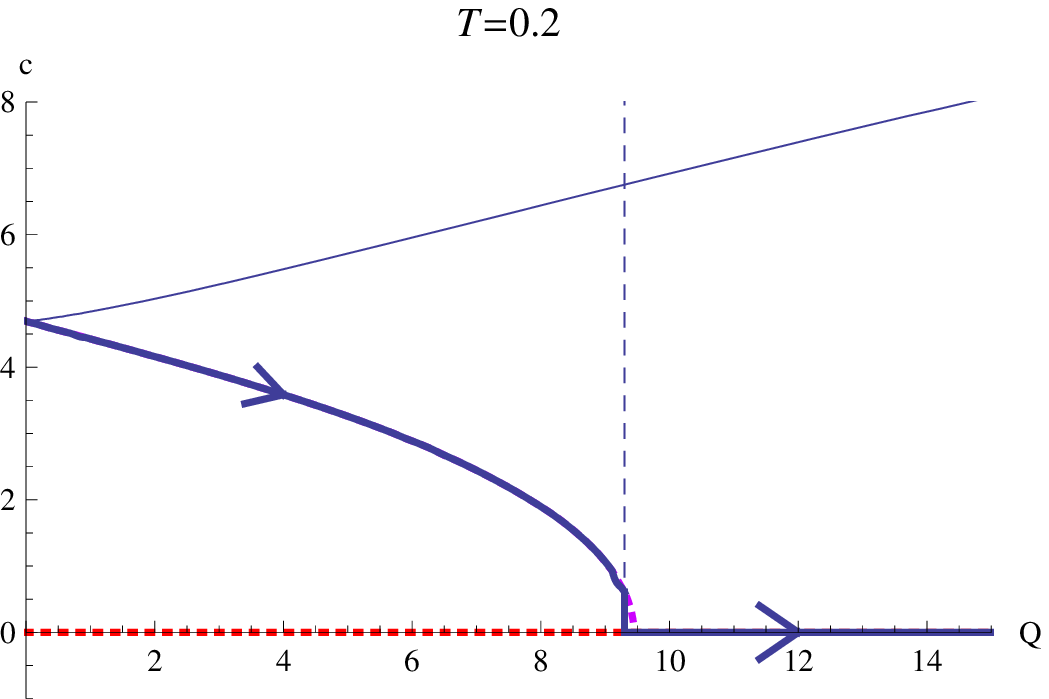}}
\hspace{1cm}
\subfigure[]{\includegraphics[angle=0,width=0.35\textwidth]{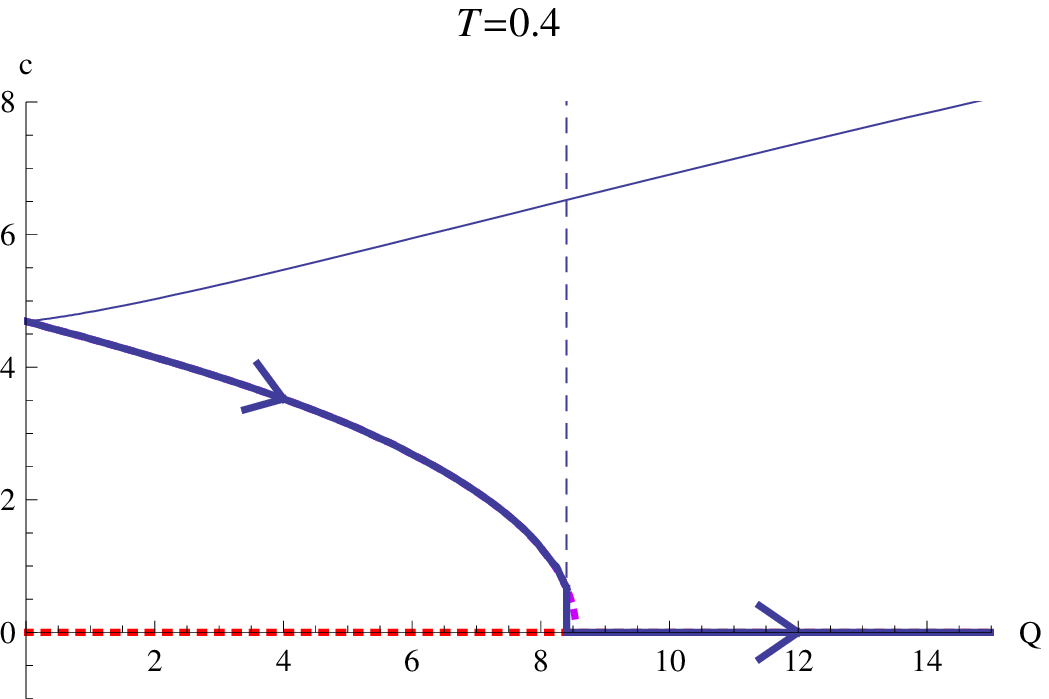}}
\caption{Density dependence of the value of chiral condensation $<\bar{q} q>$ for massless quark case with $q=15$. Thick line indicates physical phase from free energy and there is  phase transition along vertical dashed line. \label{fig:gmu01}}
\end{center}
\end{figure}

\section{Summary  and Discussion}
In this paper, we study phase structure  of a holographic QCD based on D3/D-instanton background. 
The phase transition here is for confinement/deconfiiement of quark  not the gluons. 
While the Hawking Page transition in usual AdS/CFT correspondence describes the dynamics of 
the gluon, confinement/deconfinement phase transition of quarks are determined from the interaction between the geometry and the  compact brane dynamics. Namely it is question of existence of the baryon vertex. 
  The D3/D-instanton background have quasi-confining nature as defined in the introduction.  
    In zero quark mass case, there is  a chiral transition in quark phase. 
    The   number of D-instanton,    identified as   the expectation value of gluon condensation  played, essential  role in breaking the chiral symmetry. The phase boundary is depends on value of $q$.  
 
A few remarks are in order:
 
 The first one is on the assumption that there is no Hawking-Page transition.  
Main reason is that the geometric transition should be discussed in the Einstein frame, where 
dilaton action  is cancelled by that of  axion therefore we can not find any effect of the dilaton condensation scale 
in bulk  free energy level. Therefore it is completely the same as the D3 brane without D-instanton. 
However all the probe D-brane dynamics is affected by the presence of the dilaton factor. 
That is why we get non-trivial result. This is interesting and subtle point and in the main sections 
we just assumed that there is no Hawking-Page transition. 

Second one is about the role of the Chern-Simons term. 
In ref. \cite{Ghoroku}, the authors introduced Chern-Simons term such that 
it cancel out the dilaton effect of the brane embedding dynamics. 
However, we could not find solution with such behavior. We found that the Chern-Simons term  is a total derivative 
so that it can contribute   to the charge but 
not to the equation of motion. This allows us the non-trivial effect of the gluon condensation on the   embedding of  D7 brane
as well as on that of D5. 

 The third one is that due to the probe nature of the embedding dynamics, 
 our calculation is not trusted in extreme high density regime. One need to take care of the backreaction of the geometry. 

The 4th is the Euclidean nature and duality in such background.   
For the Euclidean configuration, there is no state/geometry correspondence. 
However,  gauge gravity duality is still there.
The correspondence of  D-instanton in type IIB and and   Yang-Mill theory   instanton was discussed in  
\cite{Bianchi:1998nk} as well as many other papers. 
The relation of AdS/CFT and multi-instanton gauge theory is considered 
in well known works of Dorey  et.al. \cite{Dorey:1999pd}.  
So the gauge theory dual of the uniform distribution of such D-instanton is not hard to imagine
and it should be the Yang-Mill theory with uniform distribution of instanton charge.
 
The 5th  is the Infra-red singularity. The background has the IR singularity since 
the dilaton factor has log singularity at the horizon. Usually 
one need to prevent any D-brane probe to approach such singular region. 
Also this makes the bulk action divergent and we would need IR cut off.  
However, in our background, 
both the bulk gravity action as well as the DBI action are regular at the horizon.    There is no divergence. 
There are   reasons for this.  
i)  The finite temperature version 
 has a regular horizon and the essential singularity  is hidden in the horizon. 
ii) The finiteness of the bulk action is partly  due to the supersymmetric construction of the
 action and the ansatz eq.(3.1) by which the action of dilaton  is cancelled precisely by  that of axion. 
iii)The finiteness of the probe brane action is due to detailed structure of DBI action. Looking at eq. (4.7),  the log divergence of dilaton factor $e^\Phi$ is cancelled by the presence of the $\omega_-$ factor, 
 which has a zero.  The thermodynamics is based on the calculation of the DBI action for the actual configuration of the 
Speaking more physically, the background act net repulsive force to the D-brane which 
perform the dynamic censorship and gives finite values of DBI action.   
 
 In the background with  $q=0$, the meson spectral function does not have interesting feature. 
 But due to the repulsive nature of the force acting on the probe brane, there is some reason 
 to expect non-trivial spectral behavior in this case.   It would be very interesting to calculate the  meson spectrum in the chiral symmetry broken quark phase. 
  
  \section*{Acknowledgments}
This work was supported by the National Research Foundation of Korea(NRF) grant funded by the Korea government(MEST) through the Center for Quantum Spacetime(CQUeST) of Sogang University with grant number 2005-0049409. The work of YS and SJS was supported by Mid-career Researcher Program through NRF grant(No. 2010-0008456). The work of MK  was supported in part by KRF-2007-313- C00150, WCU Grant No. R32-
2008-000-101300.

\appendix
\section{Chern-Simons term}\label{CS}
We will put probe D7 brane, then 8-form field from hodge dual of axion can interact with D7 brane world volume. We check it in zero temperature limit and extend to finite temperature case.\par 
In the limit $T\rightarrow 0$, (\ref{10dmetric}) becomes
\bea
ds &=& e^{\Phi/2}\left\{\frac{r^2}{R^2}\left(-dt^2 +d\vec{x}^2\right) +\frac{R^2}{r^2}\left(dr^2 +r^2 d\Omega_5^2\right)\right\}\cr
e^{\Phi}&=&1+\frac{q}{r^4},~~~~\chi=-e^{-\Phi}+\chi_{\infty}.
\eea
We are interested in interaction between dual field of axion and probe D7 brane, we change the metric into the direction along and perpendicular to D7 brane,
\be
ds = e^{\Phi/2}\left\{\frac{r^2}{R^2}\left(-dt^2 +d\vec{x}^2\right) +\frac{R^2}{r^2}\left(d\rho^2 +\rho^2 d\Omega_3^2 + dy^2 +y^2 d\phi^2\right)\right\},
\ee
where $r^2 =\rho^2 +y^2$. To get dual field, we introduce vielbein 
\bea
&&e^{{\tilde t}} = e^{\Phi/4}\frac{r}{R} {\rm d}t,~~~e^{{\tilde x}_i}=e^{\Phi/4}\frac{r}{R} {\rm d}x_i,~~~e^{{\tilde \rho}}=e^{\Phi/4}\frac{R}{r} {\rm d}\rho,\cr\cr
&&e^{{\tilde \Omega}_3}= e^{\Phi/4}\frac{R}{r}\rho {\rm d}\Omega_3,~~~e^{{\tilde y}}=e^{\Phi/4} \frac{R}{r} {\rm d}y,~~~e^{{\tilde \phi}}=e^{\Phi/4}\frac{R}{r} y {\rm d}\phi,
\eea
where tilde denotes to flat index.
The field strength for axion field is
\bea
F_{(1)}&=&{\rm d} \chi = \frac{\partial \chi}{\partial r} {\rm d} r \cr
&=& e^{-\Phi} \frac{\partial \Phi}{\partial \rho} {\rm d}\rho + e^{-\Phi}\frac{\partial \Phi}{\partial y} {\rm d} y \cr
&=& e^{-5\Phi/4} \frac{\partial \Phi}{\partial \rho}\frac{r}{R} e^{{\tilde \rho}}+ e^{-5\Phi/4} \frac{\partial \Phi}{\partial y}\frac{r}{R} e^{{\tilde y}}.
\eea
Hodge dual of this field strength is

\bea
F_{(9)}&=& \frac{1}{9!} e^{-5\Phi/4} \frac{\partial \Phi}{\partial \rho}\frac{r}{R}\epsilon^{\tilde \rho}_{{\tilde t}{\tilde x}_i {\tilde \Omega}_3 {\tilde y}{\tilde \phi}} e^{{\tilde t}}\wedge e^{{\tilde x}_i}\wedge e^{{\tilde \Omega}_3} \wedge e^{{\tilde y}} \wedge e^{{\tilde \phi}} \cr
&& +\frac{1}{9!} e^{-5\Phi/4} \frac{\partial \Phi}{\partial y}\frac{r}{R}\epsilon^{\tilde y}_{{\tilde t}{\tilde x}_i {\tilde \Omega}_3 {\tilde \rho}{\tilde \phi}} e^{{\tilde t}}\wedge e^{{\tilde x}_i}\wedge e^{{\tilde \rho}}\wedge e^{{\tilde \Omega}_3} \wedge e^{{\tilde \phi}} \cr
&=& e^{\Phi} \frac{\partial \Phi}{\partial \rho} \rho^3 y {\rm d} t\wedge {\rm d}{\vec x}\wedge {\rm d}\Omega_3 \wedge {\rm d}y \wedge {\rm d}\phi \cr
&&-e^{\Phi} \frac{\partial \Phi}{\partial y} \rho^3 y {\rm d} t\wedge {\rm d}{\vec x}\wedge {\rm d}\Omega_3 \wedge {\rm d}\rho \wedge {\rm d}\phi \cr
&=&  \frac{\partial e^{\Phi}}{\partial \rho} \rho^3 y {\rm d} t\wedge {\rm d}{\vec x}\wedge {\rm d}\Omega_3 \wedge {\rm d}y \wedge {\rm d}\phi \cr
&&- \frac{\partial e^{\Phi}}{\partial y} \rho^3 y {\rm d} t\wedge {\rm d}{\vec x}\wedge {\rm d}\Omega_3 \wedge {\rm d}\rho \wedge {\rm d}\phi ,
\eea
where we are using the convention
\be
\epsilon_{{\tilde t}{\tilde x}_i {\tilde \rho}{\tilde \Omega}_3 {\tilde y}{\tilde \phi}} = +1.
\ee
By substituting $e^{\Phi}$, we get
\bea\label{dual1}
F_{(9)}&=&-\frac{4 q \rho^4 y}{(\rho^2 +y^2)^3} {\rm d} t\wedge {\rm d}{\vec x}\wedge {\rm d}\Omega_3 \wedge {\rm d}y \wedge {\rm d}\phi \cr
&&+\frac{4 q \rho^3 y^2}{(\rho^2 +y^2)^3} {\rm d} t\wedge {\rm d}{\vec x}\wedge {\rm d}\Omega_3 \wedge {\rm d}\rho \wedge {\rm d}\phi .
\eea
Now, we want to get 8-form potential such that 
\be\label{nineform} 
F_{(9)}={\rm d} C_{(8)}.
\ee
Assuming
\bea
C_{(8)}&=&f(\rho, y, \phi) {\rm d} t\wedge {\rm d}{\vec x}\wedge {\rm d}\Omega_3 \wedge {\rm d}y \cr
&&+g(\rho, y, \phi) {\rm d} t\wedge {\rm d}{\vec x}\wedge {\rm d}\Omega_3 \wedge {\rm d}\rho,
\eea
we get 
\bea
{\rm d}C_{(8)}&=& \frac{\partial f}{\partial \phi}  {\rm d} t\wedge {\rm d}{\vec x}\wedge {\rm d}\Omega_3 \wedge {\rm d}y \wedge {\rm d}\phi \cr
&& - \frac{\partial f}{\partial \rho} {\rm d} t\wedge {\rm d}{\vec x}\wedge {\rm d}\Omega_3 \wedge {\rm d}\rho \wedge {\rm d}y \cr
&& + \frac{\partial g}{\partial \phi} {\rm d} t\wedge {\rm d}{\vec x}\wedge {\rm d}\Omega_3 \wedge {\rm d}\rho \wedge {\rm d}\phi \cr
&& + \frac{\partial g}{\partial y}{\rm d} t\wedge {\rm d}{\vec x}\wedge {\rm d}\Omega_3 \wedge {\rm d}\rho \wedge {\rm d}y.
\eea
By comparing (\ref{dual1}), we get the condition for $f(\rho, y, \phi)$ and $g(\rho, y, \phi)$ as follows;
\bea\label{eqfg}
&&\frac{\partial f}{\partial \phi}=-\frac{4 q \rho^4 y}{(\rho^2 +y^2 )^3},\cr
&&\frac{\partial g}{\partial \phi}=\frac{4 q \rho^3 y^2}{(\rho^2 +y^2)^3}, \cr
&&\frac{\partial f}{\partial \rho} -\frac{\partial g}{\partial y}=0.
\eea 
By integrating $f$ and $g$ with respect to $\phi$, we get
\bea
f(\rho, y, \phi) &=&- \frac{4 q \rho^4 y}{(\rho^2 +y^2 )^3} (\phi +\phi_1)\cr
g(\rho, y, \phi) &=& \frac{4 q \rho^3 y^2}{(\rho^2 +y^2)^3}(\phi +\phi_2),
\eea
but from the last condition of (\ref{eqfg}), we get $\phi_1 =\phi_2 =\phi_0$. Finally, 8-form potential can be written
\bea\label{c8}
C_{(8)} &=& - \frac{4 q \rho^4 y}{(\rho^2 +y^2 )^3} (\phi +\phi_0) {\rm d} t\wedge {\rm d}{\vec x}\wedge {\rm d}\Omega_3 \wedge {\rm d}y \cr
&+& \frac{4 q \rho^3 y^2}{(\rho^2 +y^2)^3}(\phi +\phi_0){\rm d}t\wedge {\rm d}{\vec x}\wedge {\rm d}\Omega_3 \wedge {\rm d}\rho.
\eea
If we fix location of D7 brane along $\phi$ direction at $\phi=\phi_*$, then (\ref{c8}) becomes total derivative i.e.
\bea
C_{(8)} &=& - \frac{4 q \rho^4 y}{(\rho^2 +y^2 )^3} (\phi_* +\phi_0) {\rm d} t\wedge {\rm d}{\vec x}\wedge {\rm d}\Omega_3 \wedge {\rm d}y \cr
&+& \frac{4 q \rho^3 y^2}{(\rho^2 +y^2)^3}(\phi_* +\phi_0){\rm d}t\wedge {\rm d}{\vec x}\wedge {\rm d}\Omega_3 \wedge {\rm d}\rho \cr
&=& (\phi_* +\phi_0)\,\, {\rm d}\left[\frac{q \rho^4}{(\rho^2 +y^2)^2} {\rm d} t\wedge {\rm d}{\vec x}\wedge {\rm d}\Omega_3 \right].
\eea
so that $C_{(8)}$ is a pure gauge whose field strength is zero. 
Furthermore we can always choose $\phi_0 =-\phi_* +2\pi$, then the Chen-Simons term in (\ref{zeroTdvi}) becomes 
\bea
S_{CS} &=& \mu_{7} \int d^8 \sigma (2\pi) {\rm d}\left[\frac{q \rho^4}{(\rho^2 +y^2)^2} {\rm d} t\wedge {\rm d}{\vec x}\wedge {\rm d}\Omega_3 \right] \cr\cr
&=& (2\pi) \mu_{7} V_4 \Omega_3 \frac{q \rho^4}{(\rho^2 +y^2)^2}\Big|_{\rho=\infty} \cr\cr
&=& q (2\pi)\mu_{7} V_4 \Omega_3 \cr\cr
&=& \frac1{2}{N_{D(-1)}}.
\eea
It is nothing but the half of D-instanton number calculated in \cite{Liu:1999fc}. 
  This is completely 
satisfactory since   D7 brane world volume can  captures the flux of the upper hemi-sphere.
Since the 
Chern-Simons action is   locally a total derivative term, it does not contribute to the equation of motion.  

\section{Hamiltonian density of baryon vertex}\label{BV}
We start from the action for D5 brane with Chern-Simons term (\ref{bary-d5}),
\bea\label{D5dbi}
S_{D5} &=& S_{DBI} +S_{CS} \cr
&=& -\mu_5 \int e^{-\Phi} \sqrt{-{\rm det}(g+2\pi \alpha'
F)}+\mu_5 \int  A_{(1)}\wedge G_{(5)} \cr\cr 
&=& \t_5 \int dt d\theta \sin^4\theta 
\left[-\sqrt{e^{\Phi} \frac{\o_-^2}{\o_+} (\xi^2 +\xi'^2)-\tilde{F}^2} +4 \tilde{A}_t
\right] \cr\cr 
&=& \int dt d\theta{\cal L}_{D5}, 
\eea 
where 
\be
\t_5 = \mu_5 \O_4 R^{4}r_T,~~~~~~
\tilde{F} = 2\pi \a' F_{t\theta},~~~~~~ \tilde{A}_t= 2\pi\a'A_t.
\ee
We denote ${\cal L}_{D5}$ to Lagrangian density.
The displacement can be obtained by derivative of Lagrangian density with respect to $A_t'$,
\bea\label{DF}
D(\theta) &\equiv&  -\frac{\partial {\cal L}_{D5}}{\partial A_t'} \cr
&=& -2\pi \a' \t_5 \frac{\sin^4 \th\tilde{F}}{\sqrt{e^{\Phi} \frac{\o_-^2}{\o_+} (\xi^2 +\xi'^2)-\tilde{F}^2}}.
\eea
Then the equation of motion for gauge filed can be written as
\be\label{eqD}
\partial_\th \hat{D}(\th) =-4 \sin^4\th,
\ee
where $\hat{D}(\th)\equiv \frac{D(\th)}{2\pi\a'\t_5}$. This equation plays a constraint in the action, then the action of D5 brane becomes,
\bea
S_{D5} &=& S_{DBI} +\t_5 \int dt d\th 4 \sin^4 \th \tilde{A}_t \cr
&=& S_{DBI} - \t_5 \int dt d\th (\partial_{\th}\hat{D}(\th)) \tilde{A}_t \cr
&=& S_{DBI} - \t_5 \int dt d\th \hat{D} \tilde{F},
\eea
where we take integration by part in last procedure. This is nothing but the Legendre transformation of DBI action of D5 brane. After substituting (\ref{DF}) in the the action, we define 'Hamiltonian density'(\ref{Hd5}) as follows
\bea\label{hamd5}
S_{D5}&=& -\tau_5 \int dt d\th
\sqrt{\frac{e^{\Phi}}{2}\frac{\o_-^2}{\o_+}\left(\xi'^2 +\xi^2\right)} \sqrt{\hat{D}(\th)^2 +\sin^8
\th}\cr
&\equiv& -\int dt  d\th {\cal H}_{D5}.
\eea
The definition of Hamiltonian density is consistent with one of probe D7 brane and integration of this with on-shell solution gives free energy of D5 brane. Next, by integrating (\ref{eqD}), we get
\be\label{solD}
\hat{D}(\th)=\frac{3}{2}(\nu\pi -\th)+\frac{3}{2}\sin\th\cos\th +\sin^3 \th \cos\th,
\ee
where the integration constant $\nu$ determines the number of fundamental strings on each pole, i. e. $\nu N_c$ strings are attached to the south pole and $(1-\nu)N_c$ strings to north pole of D5 brane. Here, we assume that all the fundamental strings are attached on the north pole, we set $\nu =0$.

\section{Force balance condition}\label{FBC}

In this section, we derive force balance condition (\ref{fbc}). In our mode, the end points of fundamental strings play role of source of $U(1)$ gauge field. Due to the tension of fundamental string, there can exist cusps on probe brane world volume. By calculating force at the cusp of each brane, we can estimate the behavior of probe branes. \par
The force at the cusp can be obtained by taking small variation of 'on-shell' free energy with respect to $U_c$
\be
F_{c} =\frac{\delta {\cal F}_{\rm on-shell}}{\delta U_c},
\ee
where $U_c$ is position of cusp and the free energy is integration of Hamiltonian density of probe brane
\be
{\cal F} =\int d\r {\cal H}_{\rm on-shell}.
\ee
The Hamiltonian density is a function of $U$ and $U'$, we can write the variation as follows;
\bea
\delta {\cal H}_{\rm on-shell} &=& \delta {\cal H}(U, U' ;\rho)_{\rm on-shell} \cr\cr
&=& \frac{\partial {\cal H}_{\rm on-shell}}{\partial U}\delta U +\frac{\partial {\cal H}_{\rm on-shell}}{\partial U'}\delta U' \cr\cr
&=& \frac{\partial {\cal H}_{\rm on-shell}}{\partial U}\delta U + \frac{d}{d \rho}\left[\frac{\partial{\cal H}_{\rm on-shell}}{\partial U'} \delta U\right] -\frac{d}{d\rho}\left(\frac{\partial {\cal H}_{\rm on-shell}}{\partial U'}\right) \delta U\cr\cr
&=&\frac{d}{d \rho}\left[\frac{\partial{\cal H}_{\rm on-shell}}{\partial U'} \delta U\right] +\left[\frac{\partial {\cal H}_{\rm on-shell}}{\partial U}-\frac{d}{d\rho}\left(\frac{\partial {\cal H}_{\rm on-shell}}{\partial U'}\right) \right]\delta U\cr\cr
&=&\frac{d}{d \rho}\left[\frac{\partial{\cal H}_{\rm on-shell}}{\partial U'} \delta U\right].
\eea
Finally, the force at the cusp is
\bea\label{Fc}
F_c &=& \int d\rho \frac{\delta{\cal H}_{\rm on-shell}}{\delta U_c} \cr\cr
&=& \int d \rho  \frac{d}{d \rho}\left[\frac{\partial{\cal H}_{\rm on-shell}}{\partial U'} \frac{\delta U}{\delta U_c} \right]\cr\cr
&=& \frac{\partial {\cal H}_{\rm on-shell}}{\partial U'}\Bigg|_{U=Uc}.
\eea

The force at the cusp of single $D5$ brane can be calculated from (\ref{Fc}),
\be
F_{D5} = N_c T_{F1} \sqrt{\frac{e^{\Phi}}{2}\frac{\o_-^2}{\o_+}}\frac{\xi'}{\sqrt{\xi^2 +\xi'^2}}\Bigg|_{\xi =\xi_c},
\ee
where $T_{F1}$ is tension of fundamental string $1/2\pi \a'$ and $\xi_c$ is position of cusp of $D5$ brane. With a same manner, we can get force at the cusp of probe $D7$ brane,
\be
F_{D7} = T_{F1} \sqrt{\frac{e^{\Phi}}{2}\frac{\o_-^2}{\o_+}} \frac{Q \dot{y}}{\sqrt{1+\dot{y}^2}}\Bigg|{y=y_c},
\ee 
where $y_c$ is position of $D7$ brane at the cusp($\rho=0$) and $Q$ is total number of $U(1)$ source. The force at the cusp of $D5$ and $D7$ is always smaller than the fore of fundamental strings. Then strings pull each brane until the length of string becomes zero and force between two branes is balanced.  We can get condition of slope of $D7$ brane from the force balance condition.
\bea
0 &=& F_{D7}(Q) + N_B F_{D5} \cr
&=& F_{D7} (Q) + \frac{Q}{N_c} F_{D5} \cr\cr
&&\rightarrow \cr\cr
\dot{y_c}&=&\frac{\xi_c'}{\xi_c}.
\eea
\newpage

\newcommand{\J}[4]{{ #1} {\bf #2} #4 (#3)}
\newcommand{\andJ}[3]{{\bf #1} (#2) #3}
\newcommand{\AP}{Ann.\ Phys.\ (N.Y.)}
\newcommand{\MPL}{Mod.\ Phys.\ Lett.}
\newcommand{\NP}{Nucl.\ Phys.}
\newcommand{\PL}{Phys.\ Lett.}
\newcommand{\PR}{Phys.\ Rev.}
\newcommand{\PRL}{Phys.\ Rev.\ Lett.}
\newcommand{\ATMP}{Adv.\ Theor.\ Math.\ Phys.}
\newcommand{\JHEP}{JHEP}
\newcommand{\IJMP}{Int.\ J.\ Mod.\ Phys.}
\newcommand{\JETPL}{JETP\ Lett.}
\newcommand{\SJNP}{Sov.\ J.\ Nuc.\ Phys.}
\newcommand{\PTP}{Prog.\ Theor.\ Phys.}



\end{document}